\documentclass[aip,amsmath,amssymb,reprint,floatfix]{revtex4-2}

\usepackage{graphicx}
\usepackage{dcolumn}
\usepackage{bm}
\usepackage[utf8]{inputenc}
\usepackage[T1]{fontenc}
\usepackage{mathptmx}
\usepackage[linesnumbered,ruled,vlined,commentsnumbered]{algorithm2e}
\usepackage{amsmath}
\usepackage{amssymb}
\usepackage{xcolor}
\usepackage{soul}
\usepackage{hyperref}
\hypersetup{
	colorlinks=true,
	linkcolor=blue,
	filecolor=magenta,      
	urlcolor=blue,
	citecolor=blue
}

\definecolor{com}{HTML}{b103fc}

\begin{document}
	\preprint{AIP/123-QED}
	\title{Singlet exciton dynamics of perylene diimide and tetracene based hetero/homogeneous substrates via an \textit{ab initio} kinetic Monte Carlo model}
	\author{A. Manian}
	\email{anjay.manian2@rmit.edu.au}
	\altaffiliation{These authors contributed equally.}
	\affiliation{ARC Centre of Excellence in Exciton Science, School of Science, RMIT Univeristy, Melbourne, 3000, Australia.}
	\author{F. Campaioli}
	\email{francesco.campaioli@rmit.edu.au}
	\altaffiliation{These authors contributed equally.}
	\affiliation{ARC Centre of Excellence in Exciton Science, School of Science, RMIT Univeristy, Melbourne, 3000, Australia.}
	\author{I. Lyskov}
	\affiliation{ARC Centre of Excellence in Exciton Science, School of Science, RMIT Univeristy, Melbourne, 3000, Australia.}
	\author{J. H. Cole}
	\affiliation{ARC Centre of Excellence in Exciton Science, School of Science, RMIT Univeristy, Melbourne, 3000, Australia.}
	\author{S. P. Russo}
	\email{salvy.russo@rmit.edu.au}
	\affiliation{ARC Centre of Excellence in Exciton Science, School of Science, RMIT Univeristy, Melbourne, 3000, Australia.}
	\date{\today}
	
	\begin{abstract}
		Luminescent solar concentrators (LSCs) are devices that trap a portion of the solar spectrum and funnel it towards photon harvesting devices. The modelling of LSCs at a quantum chemical level however, remains a challenge due to the complexity of exciton and photon dynamic modelling. This study examines singlet exciton dynamics occurring within a typical LSC device. To do this, we use a rejection-free kinetic Monte Carlo method to predict diffusion lengths, diffusion coefficients, substrate anisotropy, and average exciton lifetimes of perylene diimide (PDI) and tetracene based substrates in the low concentration scheme. \textit{Ab initio} rate constants are computed using time-dependant density functional theory based methods. PDI type substrates are observed to display enhanced singlet exciton transport properties when compared to tetracene. Simulations show that substrates with dipole-aligned chromophores are characterised by anisotropic exciton diffusion, with slightly improved transport properties. Finally, a PDI-tetracene substrate is simulated for both disordered and dipole-aligned chromophore configurations. In this multi-dopant substrate transport is predominantly mediated by PDI due to the asymmetry in the transport rates between the two dyes considered. We conclude discussing the properties of multi-dopant substrates and how they can impact the design of next generation LSCs.
	\end{abstract}
	
	\maketitle
	
	\section{Introduction}
	\noindent
	
	Luminescent Solar Concentrators (LSCs) are devices which use the concept of exciton hopping and total internal reflection to trap light from a large surface area, and guide photons to photovoltaic (PV) cells of a smaller surface area\cite{emaaemilsc,tyolscr,tesctmdflsd,alscrtswagui}. Here, total internal reflection allows for transport of energy across macroscopic scales, while exciton dynamics mediate the efficient transfer of energy between molecular bodies, in addition to the recycling of otherwise unusable excitions via multi-exciton interactions\cite{Eersel2015,nmosfiams,mcwatkfdoma,taaottiopsmkcmam,oaeosateditc}. LSCs are obtained by embedding chromophores in a transparent matrix substrate with a large refractive index. When a photon enters the LSC, it is absorbed by a chromophore. Then, upon photon emission, such as fluorescence, the photon is guided within the matrix due to a condition of total internal reflection\cite{hfmipdeocopp,lilscu}. Light is thus concentrated within the matrix, and is funnelled towards the edges, where PV cells are fixed to absorb each photon\cite{emaaemilsc}.
	
	
	While the idea at the core of this strategy is simple, a few hurdles complicate the design of effective LSCs. On one hand, a large concentration of chromophores within the substrate is required such that photons will always interact with the device. However within such an amorphous matrix, the chromophores tend to aggregate and clump together, creating undesired sources of luminescent quenching\cite{FRETeploPDIbcmaai,aceoaieqmfpdi}, where quenching refers to processes competing with radiative pathways. On the other hand, if the concentration of chromophores is too low, then the majority of photons are lost as they pass through the LSC\cite{hfmipdeocopp}. Other sources of loss include photons escaping the device\cite{lilscu} upon fluorescence, and the intrinsic photoluminescence quantum yield (PLQY) loss, i.e., the probability of fluorescence of the host chromophore. Previously, we have investigated the parameters behind excitonic quenching\cite{FRETeploPDIbcmaai,aceoaieqmfpdi,aceoaieqmfpdi,NRpaper}, providing us a strong foundation with which to consider the processes within a given device. 
	
	Modelling of an LSC device at a quantum level can be split into two categories. The first concerns modelling of exciton dynamics, i.e., how long the exciton takes to dissociate and return to a photon. The second concerns modelling of the photon dynamics, i.e., when an exciton has dissociated into a photon, how that energy packet propagates through the substrate, whether it is reabsorbed by another chromophore, by the PV, or escapes. The former requires an understanding of the quantum chemistry and corresponding rate constants of the numerous de-excitation pathways, while the latter requires tracking of photons through the substrate, or ray tracing\cite{mcrtsolscfbip,tsaaolsBMPLSCb3dmcrtm}.
	
	The dynamics of both interacting and non-interacting excitons have been simulated using kinetic Monte Carlo (KMC) methods in a variety of studies, providing insights on the fundamental features of exciton transport as well as practical guidelines for the design of optoelectronics and PV applications~\cite{Athanasopoulos2007,Kimber2012,Yost2012,Akselrod2014a,Eersel2015,Xie2016,Kaiser2018,Ligthart2018,Saxena2020}. Engels in particular has studied exciton diffusion and charge-carrier mobilities using Marcus theory\cite{marcus,fpcoaccmiosc,iourpaamrfiefipbos}. However, it was noted that diffusion properties calculated using Marcus theory were drastically underestimated with respect to experiment and other methods\cite{sefiocbomtr}, despite being qualitatively correct with respect to trends. Further, the Monte Carlo approach allows simulation of transport properties within a time dependent framework. As such, a method involving computation of the full spectral density \textit{within} a time dependent framework may allow for accurate results comparable with experiment. 
	
	While modelling of the photons' dynamics in the substrate has been studied rigorously\cite{alscrtswagui,solsceopafa,mcsolpilscbosn,pomkrtsiqdsc}, a treatment at the atomic scale is very complex; we will therefore address this problem in a future study. In this paper we focus on exciton migration through a medium such as an LSC. To do so, we use a KMC method based on transition rates that we obtain from quantum chemical calculations~\cite{NRpaper}. KMC methods have been used in surface based simulations\cite{apgtskmcs}, and exciton diffusion/hopping studies\cite{edionfamcsoteotad}, making it an ideal tool for our purpose. Herein, we simulate exciton diffusion and dissociation in homogeneous and heterogeneous Perylene Diimide (PDI) and Tetracene chromophores in a Toluene medium. PDIs\cite{aceoaieqmfpdi,hfmipdeocopp} are important in the field of organic light emitting diodes (OLEDs) and LSCs, typically displaying large PLQYs, while Tetracene\cite{sfitaesa,tdosfictaca,esdisamttrosraef} has been well studied as a host for singlet fission type applications. We investigate transport properties such as the average diffusion lengths, diffusion coefficient, and the anisotropy profile of the exciton migration across disordered chromophore configurations with both random and forcibly aligned transition dipoles. For convenience, we refer to these two types of systems as disordered configurations and dipole-aligned configurations respectively. We also study the frequency of different dissociation pathways within favoured dissociation sites, and the average exciton lifetimes, which can provide valuable guidelines for LSCs design. 
	
	\section{Theory \& Implementation}
	\subsection{Transport Model}
	Exciton transport, often referred to as exciton diffusion, is one of the key processes behind the operation of organic optoelectronic devices\cite{muoosc,dkmcfsdecaetidm,edionfamcsoteotad}. In addition to the exciton transport, we also need to model the processes that result in loss of the electron-hole quasiparticle, such as charge recombination, internal conversion (IC) and inter-system crossing (ISC). These latter two are relaxation processes mediated by heat, in which heat is exchanged with the environment when spin is and is not conserved~\cite{NRpaper,aceoaieqmfpdi} respectively.
	
	In essence, exciton transport corresponds to the the migration of an energy packet from some molecule A to another molecule B, simply shown as:
	\begin{equation}
		{}^1A^\star+{}^1B\rightarrow{}^1A+{}^1B^\star.
	\end{equation}
	Here, an exciton residing on chromophore A is transferred, or hops, to chromophore B. Exciton hopping can be modelled within both Coulombic and Exchange picture, using either F\"{o}rster theory\cite{forster} or Dexter theory\cite{dexter} respectively. In this work, we only consider F\"{o}rster type hopping, which is modelled using rigorous quantum chemical calculations, as discussed in the next section.
	
	\begin{figure}[b!]
		\centering
		\hrulefill
		\vspace{10pt}
		\includegraphics[width=\linewidth]{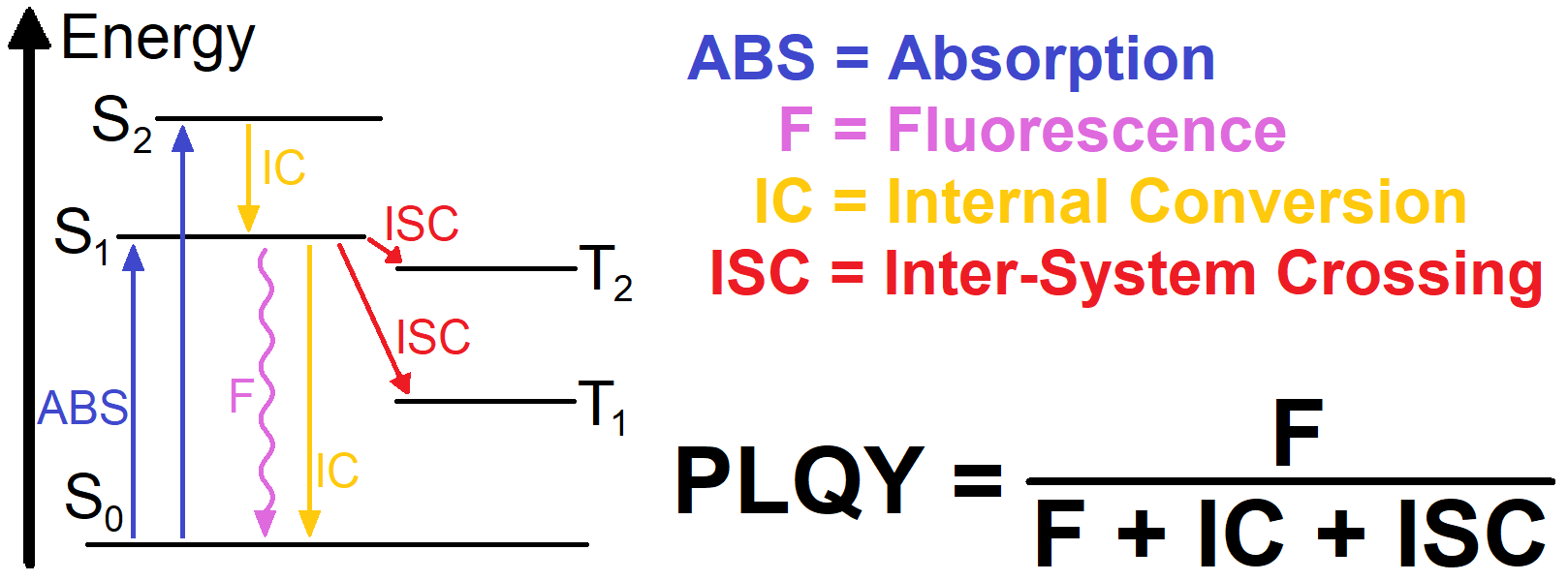}
		\caption{Energy diagram illustrating how the system excites and relaxes. Here, an incident photon excited the ground state chromophore to an excited state, be it the first or second singlet excited state. Following relaxation to the ground vibrational state of first singlet electronic excited state, the energy can transition between nearby electronic excited states. }
		\label{imagePLQY}
		\hrulefill
	\end{figure}
	
	Two chromophores that undergo energy transfer are assumed to be in resonance, where molecule A donates an exciton to an accepting molecule B, which is known as F\"{o}rster Resonance Energy Transfer (FRET). The transition rate of a FRET mechanism can be modelled using Fermi's Golden Rule, where the matrix element can be shown using the simple dipole approximation\cite{veoroeetattd}, given as:
	\begin{equation}
		\left|\left<f\left|H^{\left(1\right)}\right|i\right>\right|=\frac{1}{4\pi\epsilon_0}\frac{\kappa\left|\mu_D\right|\left|\mu_A\right|}{n^2r^3},
		\label{equationQMVDA}
	\end{equation}
	where $\mu_D$ ($\mu_A$) is the transition dipole moment of a given donor (acceptor) chromophore, $\epsilon_0$ is the permittivity of free space, $\kappa$ is the dipole angular orientation factor, given in Eq.~\eqref{fretkappa}, $n$ is the refractive index of the medium, and $r$ is the interchromophore separation. The density of states across the energy continuum $\rho$ can be expressed as the integral of the normalised absorption band of the acceptor molecule and emission band of the donor, $\textrm{Abs}\left(\omega\right)$ and $\textrm{Emi}\left(\omega\right)$ respectively, weighted by the PLQY of the donor in the absence of the acceptor, $Q_D$:
	\begin{equation}
		\rho = Q_D\int \textrm{Abs}\left(\omega\right)\textrm{Emi}\left(\omega\right)d\omega.	\label{equationSpectralOverlap}
	\end{equation}
	The angular orientation factor $\kappa$ can be expressed as:
	\begin{equation}	            
		\kappa=\hat{\mu}_A\cdot\hat{\mu}_D-3\left(\hat{\mu}_D\cdot\hat{R}\right)\left(\hat{\mu}_A\cdot\hat{R}\right),
		\label{fretkappa}
	\end{equation}
	where $\hat{R}$ is the normalized inter-chromophore separation, and where all three vectors are unit vectors. We then substitute both Equation~\ref{equationQMVDA} and Equation~\ref{equationSpectralOverlap} into Fermi's Golden Rule to yield the following:
	\begin{equation}
		\begin{split}
			k_{FRET}=&\frac{2\pi}{\hbar}\frac{1}{\left(4\pi\epsilon_0\right)^2}\frac{\kappa^2\left|\mu_D\right|^2\left|\mu_A\right|^2}{n^4r^6} \times \\
			& \times Q_D\int \textrm{Abs}\left(\omega\right)\textrm{Emi}\left(\omega\right)d\omega.
			\label{equationFRET}
		\end{split}
	\end{equation}
	All terms within the integrand are often collected together and termed the spectral overlap.
	
	When we consider excitonic quenching, we examine the probability of radiative decay upon charge recombination. This comes back to the quantum yield $Q_D$ of the chromophore. As per Kasha's Rule\cite{kasha}, fluorescence will occur from the first singlet excited state. Therefore, the quantum yield of fluorescence is a ratio of radiative decay rates against the combined rate of all possible mechanisms, as shown in Figure~\ref{imagePLQY}. In the case of an isolated monomer, the likely competing mechanisms are IC and ISC. Each process is dependant on different photophysical coupling terms: fluorescence on the transition dipole moment, IC on vibronic coupling between singlet states, and ISC the spin-orbit coupling between singlet and triplet states. These rate constants can be very difficult to compute, however we have recently developed a methodology to do so at the Time-Dependant Density Functional Theory (TDDFT) level of theory\cite{NRpaper}. 
	
	\subsection{The KMC method}
	KMC is a numerical method to simulate the dynamics of a process, based on known transition rates between the states of a considered system. KMC methods are widely adopted in many areas of Physics and Chemistry due to their computational efficiency, and can be used to simulated both equilibrium and non-equilibrium processes~\cite{Voter2007,Stamatakis2011,Walker2015,Ostroverkhova2016,Oberhofer2017,Lin2018,Jorgensen2018,Andersen2019,Rego2020,sbrtqjiasse}. They provide an ideal approach to simulate FRET and Dexter exciton transport processes in disordered systems, which are known to correspond to a classical random walk over an ensemble of sites~\cite{Kranz2016a,dkmcfsdecaetidm}.
	Moreover, when the occurrence of multi-exciton interactions processes, such as triplet fusion and singlet fission, can be neglected, each exciton can be treated independently, turning the KMC simulation into an embarrassingly parallel computational problem.
	
	In this work we adopt a rejection-free KMC method to simulate the dynamics of an ensemble of non-interacting excitons that undergo FRET-mediated transport, fluorescence, IC or ISC. The molecular geometries and rates associated with these processes were computed using a combination of DFT and DFT based multireference configuration interaction DFT/MRCI approaches\cite{MRCI_igor}. The considered systems are static arrangements of PDI and Tetracene chromophores in a substrate. This is done to simulate the transport of excitons in a solid, where chromophores' position and dipole orientation do not vary during the transport process. The photo-physical properties of the substrate are based on Toluene --- despite it being a liquid solvent ---, due to its high refractive index, ideal for the requirements of total internal reflection necessary for the LSCs working mechanism. For each considered system, we simulate transport within a cubic portion of the LSCs, whose size is such that the probability of transport outside the simulation box is negligible. In all cases we consider concentrations corresponding to a low concentration solution, such that chromophores are close enough to maximise exciton transfer, but not so close such that exciton quenching pathways facilitated by molecular aggregation can be considered negligible. 
	
	It should be noted that since we do not consider photon dynamics, we therefore do not consider photon reabsorption, a process whereby exciton dissociation at site $\vec{r}_i$ results in a photon propagating through the substrate, and is reabsorbed by another site $\vec{r}_j$. Rather, simulations begin and terminate at the \textit{birth} and \textit{death} of a given exciton, due to the foreseen complexity of ray tracing at the quantum chemical level. 
	
	First, we generate the geometry of the considered systems. A number of sites, corresponding to the chromophores locations, is randomly generated to match the dyes concentration within the simulation box. The sites are uniformly distributed in the considered volume. Then, the dipole orientation of each chromophore is randomly generated, with different distributions being used for disordered and dipole-aligned materials. Once the geometry of the systems is fixed, exciton dynamics is simulated using the KMC method. A state $\mathcal{S}_t = \{t, \vec{r}_i\} $ of the process at time $t$ is given by the location (site) $\vec{r}_i$ of a singlet exciton. Exciton transport is simulated propagating the state in time $\mathcal{S}_t \to \mathcal{S'}_{t+\Delta t}$, using the routine outlined in Alg.~\ref{kmc-routine}, until a dissociation event occurs. Transport occurs via FRET from the current site $\vec{r}_i$ to a site $\vec{r}_j$ within the interaction range $r_\text{max}$, which depends on the dye and is determined as specified in Sec.~\ref{ss:quantum_chem}.
	
	\IncMargin{1em}
	\begin{algorithm}[b]
		\SetKwData{Left}{left}\SetKwData{This}{this}\SetKwData{Up}{up}
		\SetKwFunction{Union}{Union}\SetKwFunction{FindCompress}{FindCompress}
		\SetKwInOut{Input}{input}\SetKwInOut{Output}{output}
		
		\Input{The current state $\mathcal{S}_t$.}
		\Output{The propagated state $\mathcal{S}'_{t+\Delta t}$ if FRET occurs, the dissociation pathway otherwise.}
		\BlankLine
		
		Calculate the FRET transition rates $k_{FRET}(i\to j)$ from the current site $\vec{r}_i$ to any site $\vec{r}_j$ within interaction range $r_{\text{max}}$; 
		
		Determine the dissociation rates $k_r$, $k_{IC}$, and $k_{ISC}$, which depend on the chromophore species of the current site $\vec{r}_i$;
		
		Calculate the cumulative function $R_m = \sum_{m=1}^M k_m$ of all the $M$ rates involved, and the cumulant $Q = R_M$;
		
		Sample a uniform random number $u \in (0,1]$;
		
		Determine which event occurs, by choosing the rate $k_m$ such that $R_{m-1}\leq u Q \leq R_{m}$;
		
		\If{$k_m = k_{FRET}(i\to j)$}{
			FRET occurs, the new site is $\vec{r}_j$;
			
			Sample another uniform random number $u' \in (0,1]$;
			
			Determine the time-interval $\Delta t = Q^{-1} \ln(1/u')$;
			
			Return the new state $\mathcal{S}'_{t+\Delta t}$.
		}
		\Else{
			Dissociation occurs via $k_r$, $k_{IC}$ or $k_{ISC}$;
			
			Return the dissociation pathway, \textbf{end} the routine.
		}
		
		\caption{Rejection-free KMC routine}\label{kmc-routine}
	\end{algorithm}
	\DecMargin{1em}
	
	\section{Computational Details}
	\subsection{Systems generation and KMC implementation}
	Using the KMC method described in the previous section we simulated exciton transport for single and multi-dopant substrates for two classes of 3D systems: disordered chromophore configurations with both random and aligned transition dipoles, labelled as disordered configurations and dipole-aligned configurations respectively. The former case of disordered geometries were chosen as they most closely represented exciton diffusion between nanofibres in bulk\cite{edionfamcsoteotad}. The latter case corresponds to the scenario for which most dipoles are aligned and parallel to a given axis.  Despite of the fact that dipole alignment methods can vary greatly\cite{plfomm}, the resulting geometries allow for possible circumvention of high dye concentration loss mechanisms\cite{solsceopafa,solsceopafa,edloostfmbsrplq,oaeosateditc}. 
	
	\begin{table*}
		\begin{ruledtabular}
			\caption{\label{TableTransferProperties} Calculated rates and transfer properties for Perylene Diimide and Tetracene monomers simulated in Toluene. }
			\begin{tabular}{c|ccc|cccc|cc}
				Species & $\mu_D$ $(au)$ & $\mu_A$ $(au)$ & $J$ $(au^{-1})$ & $Q_D$ & $k_r$ $(s^{-1})$ & $k_{IC}$ $(s^{-1})$ & $k_{ISC}$ $(s^{-1})$ & $R_C$ $(nm)$ & $R_I$ $(nm)$\\
				\hline
				PDI & 4.005 & 3.836 & 46.243 & 0.99 & $1.505\times 10^8$ & $1.316\times 10^4$ & $1.025\times 10^6$ & 1.6 & 15\\
				Tetracene & 1.340 & 1.299 & 31.734 & 0.26 & $2.245\times 10^7$ & $2.707\times 10^7$ & $3.565\times 10^7$ & 1.4 & 8
			\end{tabular}
		\end{ruledtabular}
	\end{table*}
	
	To generate each arrangement, chromophore concentrations were set to $6.02\times 10^6~\mu m^{-3}$ and $8.03\times 10^6~\mu m^{-3}$ for homogeneous PDI and Tetracene systems, respectively (see Supplementary Information for further details on heterogeneous chromophore concentrations). Simulations were carried out in a cubic box with side length of $0.25~\mu m$. Random sites were uniformly generated to match the desired concentration. An iterative approach was used to ensure that inter-chromophore separations were always larger than the culling sphere radius $r_{\text{min}}$, which depends on the chromophore species, as discussed in Sec.~\ref{s:results-discussion}. This was done by repopulating the simulation box after removing the sites that were too close too each other.
	
	For disordered arrangements, dipole orientations $\hat{\mu} = (\cos\theta\sin\phi,\sin\theta\sin\phi,\cos\phi)$ were randomly generated by uniformly sampling angles $\theta \in (-\pi,\pi]$ and $\phi \in (0,\pi]$, to obtain a uniform distribution on the sphere. For the case of \textit{dipole-aligned} arrangements, the two angles were sampled uniformly in $\theta\in(-p\pi,p\pi]$ and $\phi\in((1-p)\pi/2,(1+p)\pi/2]$, with $p=0.01$, thus returning dipoles mostly aligned with the $\hat{x}$ direction. Defects were simulated by randomly changing $p\to1$, with defect probability of $1\%$.
	
	For every KMC trajectory a singlet exciton was initialised at time $t_0 = 0$~ns, on a random site $\vec{r}_i$ within a sphere of radius $2 r_{\text{max}}$ located in the centre of the simulation box. Initial states $\mathcal{S}_{t_0}$ were then propagated with the KMC routine describe in Alg.~\ref{kmc-routine} until a dissociation event occurred. For each system, 5000 trajectories were sampled.
	
	\begin{figure*}[t]
		\centering
		\includegraphics[width=\linewidth]{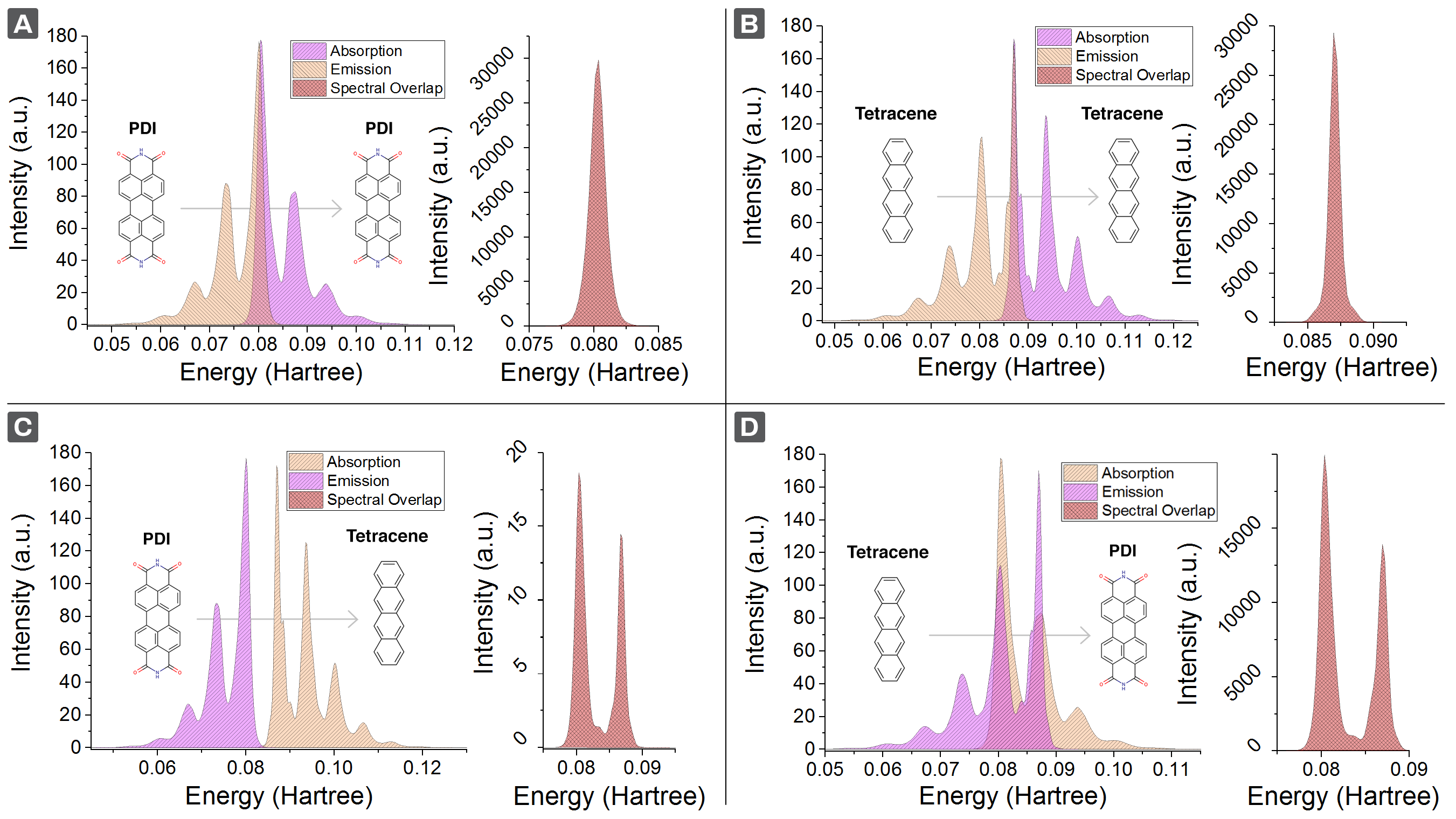}
		\caption{(\emph{Top}) Normalised absorption and emission spectra for homogeneous A$\left)\right.$ Perylene Diimide and B$\left)\right.$ Tetracene systems solvated in Toluene, in atomic units. (\emph{Bottom}) Spectral overlap schemes for heterogeneous 2 dye substrates when the exciton transport path is from C$\left)\right.$ Perylene Diimide to Tetracene, and D$\left)\right.$ Tetracene to Perylene Diimide. Paired overlap functions are computed upon combination of absorption and emission spectra, as per Equation~\ref{equationQMVDA}. }
		\label{fig:fret_rates}
		\hrulefill
	\end{figure*}
	
	\begin{figure}[t]
		\centering
		\includegraphics[width=0.95\linewidth]{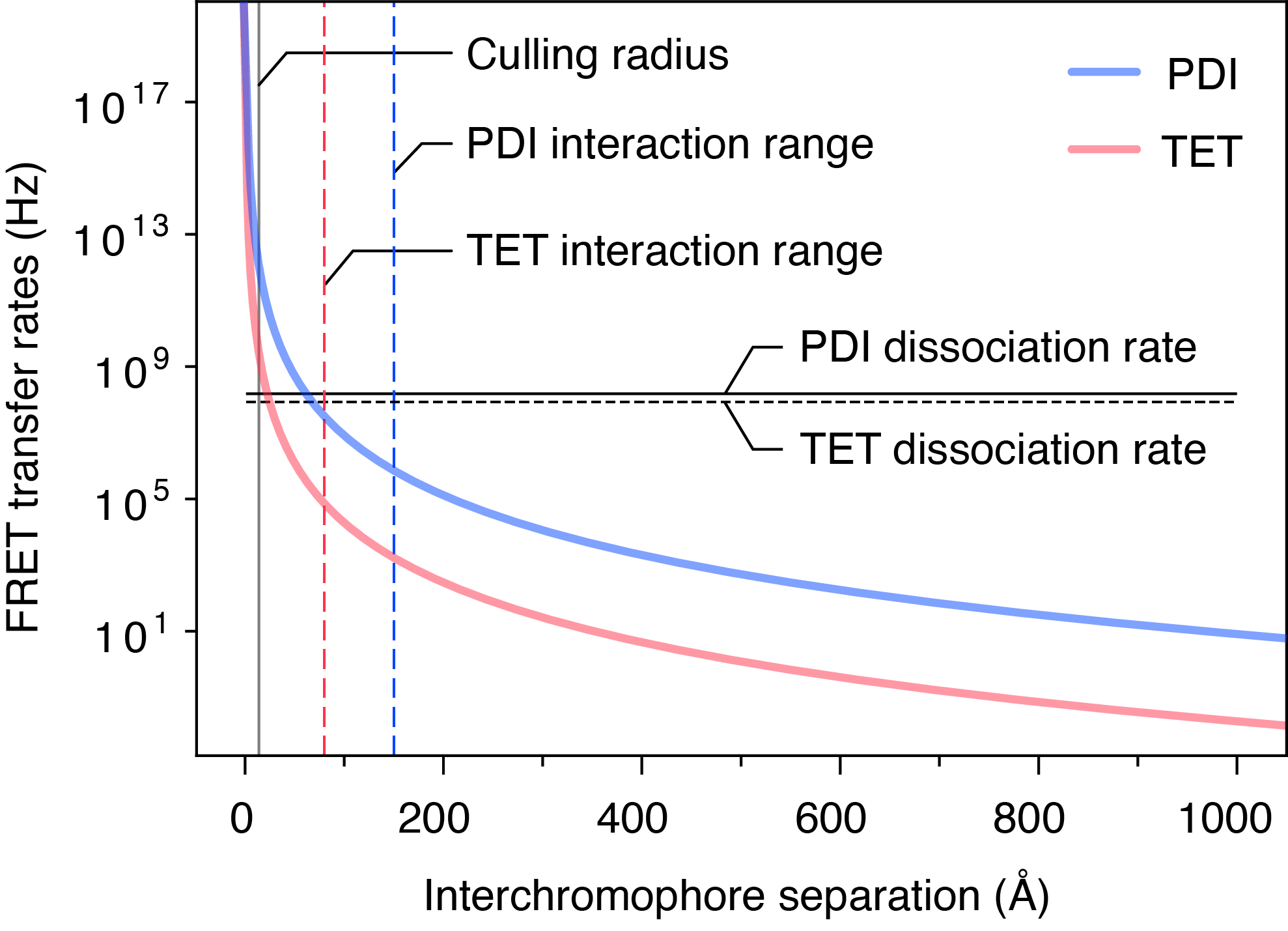}
		\caption{Excitonic transfer rate computed using F\"{o}rster theory as a function of interchromophore separation for homogeneous PDI and Tetracene (TET) substrates. 
			The horizontal lines represent the combined dissociation rates $k_{\text{diss}} = k_r +k_{IC} +k_{ISC}$ for the two dyes. Interaction ranges for PDI and Tetracene are represented by the vertical dashed lines, while the culling radii,
			equal to 16 \r{A} and 
			14 \r{A} for PDI and Tetracene, respectively, are approximately 
			represented by the solid black vertical line.}
		\label{imagePDIInteractionRadius}
		\hrulefill
	\end{figure}
	
	\subsection{Quantum Chemistry}
	Quantum chemically optimised geometries of the relevant excited states, rate constants, and photophysical properties of PDI and tetracene were computed as per our previous study\cite{NRpaper}. Here, electronic ground and excited state geometries were computed at the DFT level of theory using the Becke 3-parameter Lee–Yang–Parr exchange-correlation hybrid functional B3LYP\cite{b3lyp1} using the valence triple-zeta polarization basis set TZVP\cite{TZVP} as implemented in the \textsc{Gaussian16} software package\cite{GAUSSIAN}. A solvent of Toluene was simulated using a polarizable continuum model (PCM). The same parameters were used to compute the electronic Hessian. Single-point calculations were performed using DFT/MRCI, using the def2-TZVP basis set. The one-particle basis was computed using the Becke half-and-half Yang-Lee-Parr BHLYP exchange-correlation functional\cite{bhlyp} as implemented in the \textsc{Turbomole} software package\cite{TURBOMOLE}. Here, the PCM was employed using the \textsc{Cosmo} module\cite{COSMOiiTURBOMOLE}. Spin-orbit matrix elements were computed using the \textsc{Spock.ci} module\cite{socoDFTMRCIwf,egomefoesoo,SPOCKCI} of the DFT/MRCI platform. System temperature was set to 300 K, corresponding to standard room temperature conditions. 
	
	The culling radius was derived from the Lennard-Jones potential. Here, each atom is modelled as a sphere, with radius equal to it's collision radius dependant on it's isotropic static polarizability. Following, a box is built around these spheres, such that the volume of this structure in minimised, hugging the spheres\cite{gcitopfvdwippc,teodcoal2o3ncidbg}. The culling radius is taken as the largest dimension of this box, and is computed using the \textsc{Chemcraft} visualisation package\cite{ChemCraft}. 
	
	\section{Results \& Discussion}
	\label{s:results-discussion}
	\subsection{Quantum Chemistry}
	\label{ss:quantum_chem}
	
	Expansion of the DFT/MRCI wavefunction for PDI shows a strong $S_0\rightarrow S_1$ transition from the highest unoccupied molecular orbital (HOMO) to the lowest unoccupied molecular orbital (LUMO), agreeing well with Yang \& Jang\cite{tionfetmipdidpbatdas}. Here, absorption and emission energies were computed to be 2.36 eV and 2.09 eV respectively, with a corresponding adiabatic energy of 2.23 eV. Absorption and emission spectra, shown in Figure~\ref{fig:fret_rates}A, compares well  with experiment\cite{hfmipdeocopp}. While typically at the Franck-Condon point we expect a degree of symmetry with respect to the zero-phonon line of absorption to emission spectra, here we see a very slight shift in the density of states in the absorption spectra, with a slightly smaller secondary peak with respect to the emission bandstructure. We attribute this to the switching of intensities between two vibrational normal modes at the Franck-Condon points of the two transitions, more commonly known as the Duschinsky effect\cite{tiotesimamctFCp}. This change does not appear to affect the rate of either exciton transfer or de-excitation pathways. Tetracene also yielded a strong HOMO$\rightarrow$LUMO transition for the first singlet excited state upon DFT/MRCI wavefunction expansion. Absorption and emission energies were found to be 2.61 eV and 2.28 eV, with a resulting adiabatic energy of 2.44 eV. Similar to PDI, the spectral bandshapes compare well to literature\cite{esdisamttrosraef}, while also displaying some slight asymmetry about the 0-0 phonon line, as shown in Figure~\ref{fig:fret_rates}B.
	
	Rate constants and important quantum chemical qualities are reported in Table~\ref{TableTransferProperties}. For a homogeneous PDI exciton transfer pathway, the computed bandshapes result is a large spectral overlap of 46~$au^{-1}$. Further, donor and acceptor transition dipole moments of 4.00 au and 3.84 au were measured. For the three monomer deexciatation pathways, rate constants of $1.505\times 10^8~s^{-1}$, $1.316\times 10^4~s^{-1}$, and $1.025\times 10^6~s^{-1}$ were used for fluorescence, IC and ISC respectively, as per Ref.~\onlinecite{NRpaper}, resulting in a near-unity fluorescence quantum yield of 0.99.
	
	Tetracene on the other hand displays significantly weaker fluorescent properties. While displaying a sizeable homogeneous overlap of 32~$au^{-1}$, the donor and acceptor transition dipole moments are almost four times smaller than that of PDI, computed to be 1.34~$au$ and 1.30~$au$ respectively. The computed rate constants of $2.245\times 10^7~s^{-1}$, $2.707\times 10^7~s^{-1}$, and $3.565\times 10^7~s^{-1}$ for fluorescence, IC, and ISC respectively result in a much smaller quantum yield of 0.26. It is worth noting here that while photophysical properties computed for Tetracene in Toluene here are very similar to those computed in CycloHexane\cite{NRpaper}, smaller derivative components for the second order corrections result in a significantly smaller ISC rate constant, and therefore a larger quantum yield. 
	
	Upon comparison to experimental data reported by Burgdorff and coworkers\cite{ppotdis2htd}, we see that our predicted PLQY is slightly overestimated with respect to experiment. The fast fluorescence lifetime noted as 4.8~ns infers the stabilisation of higher lying excited states, increasing the effect of second order terms. However, Burdett and coworkers\cite{esdisamttrosraef} noted a range of lifetimes spanning 20-100~ns. This suggests that aggregation in Tetracene has a negative effect on the photophysics of the substrate, with strong aggregation slowing down the fluorescence lifetime. As such, the rates computed here can be deemed accurate with respect to a chromophore in isolation.
	
	\begin{table}[t]
		\begin{ruledtabular}
			\caption{\label{TableOverlap} Computed spectral overlap for multi-dopant substrates. Donors are given in column, while acceptors are given in row. Overlaps in units of $au^{-1}$. }
			\begin{tabular}{c|cc}
				Transfer Path $\nearrow$ & PDI & Tetracene \\
				\hline
				PDI & 46.243 & 0.056\\
				Tetracene & 59.875 & 31.734
			\end{tabular}
		\end{ruledtabular}
	\end{table}
	
	If we examine how the exciton hopping rate of an example PDI dimer system evolves as a function of the interchromophore separation, as shown in Figure~\ref{imagePDIInteractionRadius}, we can see that any separation beyond 50~nm yields rate constants smaller than $10^3~s^{-1}$. Here, all terms used for computation of the rate constant are fixed, as per Equation~\ref{equationFRET}, assuming this test system is anisotropic\cite{lakowitz} and taking the standard value of $\kappa^2=\frac{2}{3}$ for the orientation. With $r_{\text{min}}$ as a limiting factor, measured here as 16~\AA~(see Supplementary Information), we may then define a range of inter-chromophore separations in which excitonic processes dominate. Here, we note that the rate drops below $10^5~s^{-1}$ for separations larger than 15~nm. In this regime, exciton dissociation is the dominating process. We will therefore adopt this as the maximum to the interaction range $r_{\text{max}}$ for PDI chromophores. The same logic applies for a tetracene anisotropic dimer, where the exciton hopping rate drops to less than $10^5~s^{-1}$ when dimer separation increases beyond $r_{\max} = 8$~nm. Furthermore, the culling sphere radius $r_\text{min}$ for Tetracene can be taken as 14~\AA~(see Supplementary Information).
	
	In this study we also consider a novel system whereby an even mixture of the PDI and Tetracene dyes are in the same substrate. For this system, transition probabilities will not be symmetric upon the exchange of donor and acceptor, and must be calculated separately. Our results show that PDI-Tetracene displays a drastically different overlap to Tetracene-PDI, as shown in Table~\ref{TableOverlap} and Figure~\ref{fig:fret_rates}. More specifically, for the transfer of an exciton from PDI-Tetracene, a very poor overlap of 0.056~$au^{-1}$ is observed, as shown in Figure~\ref{fig:fret_rates}C. Here, the respective donor and acceptor spectra are not compatible, suggesting that PDI-PDI exciton hopping is more favourable.
	
	However, in the reverse case, where an exciton propagates via a Tetracene-PDI pathway, we see an appreciable overlap of 60~$au^{-1}$, as shown in Figure~\ref{fig:fret_rates}D. Here, the primary peaks of both respective donor and acceptor spectra overlap considerably, resulting a less intense overlap across a larger portion of the energetic continuum. This larger overlap suggests that an exciton pathway of Tetracene-PDI is in fact more favourable than homogeneous hopping, with very interesting implications on exciton transport properties for the case of mixed dyes. In this particular case, Tetracene has significantly poorer photophysical properties with respect to PDI. Therefore, in this 2-dye system, an exciton is less likely to hop from a PDI onto a Tetracene, and if it does, is likely to jump back. This asymmetry can be exploited, when trying to ensure charge recombination via a particular pathway, in this case fluorescence. 
	
	\subsection{Exciton transport properties}
	\label{ss:exciton_transport_properties}
	\begin{table*}[t]
		\caption{\label{Results_disordered}Exciton transport properties for disordered and dipole-aligned configurations. Generalised diffusion coefficient $D$ and diffusion anomaly $\alpha$ have been obtained fitting the data to the models of Eq.~\eqref{eq:general_transport} and Eq.~\eqref{eq:diffusion_anomaly}, respectively. Mean exciton lifetime $\langle \tau\rangle$ and average diffusion length  $\langle r \rangle$ have been obtained averaging over the KMC trajectories. Transport anisotropy is shown via the $D_x$, $D_y$, and $D_z$ components of $D$. The ratio of each dissociation pathways is shown for radiative (Rad), inter-conversion (IC), and inter-system crossing (ISC). See sec.~\ref{ss:exciton_transport_properties} for further details.}
		\begin{tabular}{c}
			\textbf{Disordered Configurations:} \\
			\begin{ruledtabular}
				\begin{tabular}{c|cc|cc|c|ccc}
					{Species} & $D \; (\text{nm}^2/\text{ns})$  & $\alpha$  &  $\langle \tau\rangle$ $(\text{ns})$ & $\langle r \rangle$ $(\text{nm})$ & Anisotropy $(D_x,D_y,D_z)$  & Rad (\%) & IC (\%) & ISC (\%) \\ \hline
					PDI & 2.82 & 0.94  & 6.51 & 4.76 & 0.33, 0.32, 0.34 & 99.34 & 0.64 & 0.02 \\ \hline
					TET & 0.02 & 0.84 & 11.89 & 0.46 & 0.33, 0.32, 0.34 & 26.12  & 42.82 & 31.06 \\ \hline
					PDI + TET & 0.89 & 0.94 & 7.65 & 2.78 & 0.33, 0.33, 0.33 & 83.14 & 9.80 & 7.06 \\ \hline
				\end{tabular}
			\end{ruledtabular}\\
			\vspace{-2pt}\\
			\textbf{Dipole-aligned Configurations:}\\
			\begin{ruledtabular}
				\begin{tabular}{c|cc|cc|c|ccc}
					{Species} & $D \; (\text{nm}^2/\text{ns})$  & $\alpha$  &  $\langle \tau\rangle$ $(\text{ns})$ & $\langle r \rangle$ $(\text{nm})$ & Anisotropy $(D_x,D_y,D_z)$  & Rad (\%) & IC (\%) & ISC (\%) \\ \hline
					PDI & 3.42 & 0.95  & 6.66 & 5.23 & 0.51, 0.24, 0.25 & 99.34 & 0.64 & 0.02 \\ \hline
					TET & 0.03 & 0.81 & 11.58 & 0.5 & 0.50, 0.24, 0.26 & 26.04  & 41.58 & 32.38 \\ \hline
					PDI + TET & 1.10 & 0.97 & 7.56 & 3.15 & 0.49, 0.25, 0.26 & 85.76 & 8.49 & 5.74 \\ \hline
				\end{tabular}
			\end{ruledtabular}
		\end{tabular}
	\end{table*}
	
	\begin{figure*}[t]
		\centering	\includegraphics[width=\linewidth]{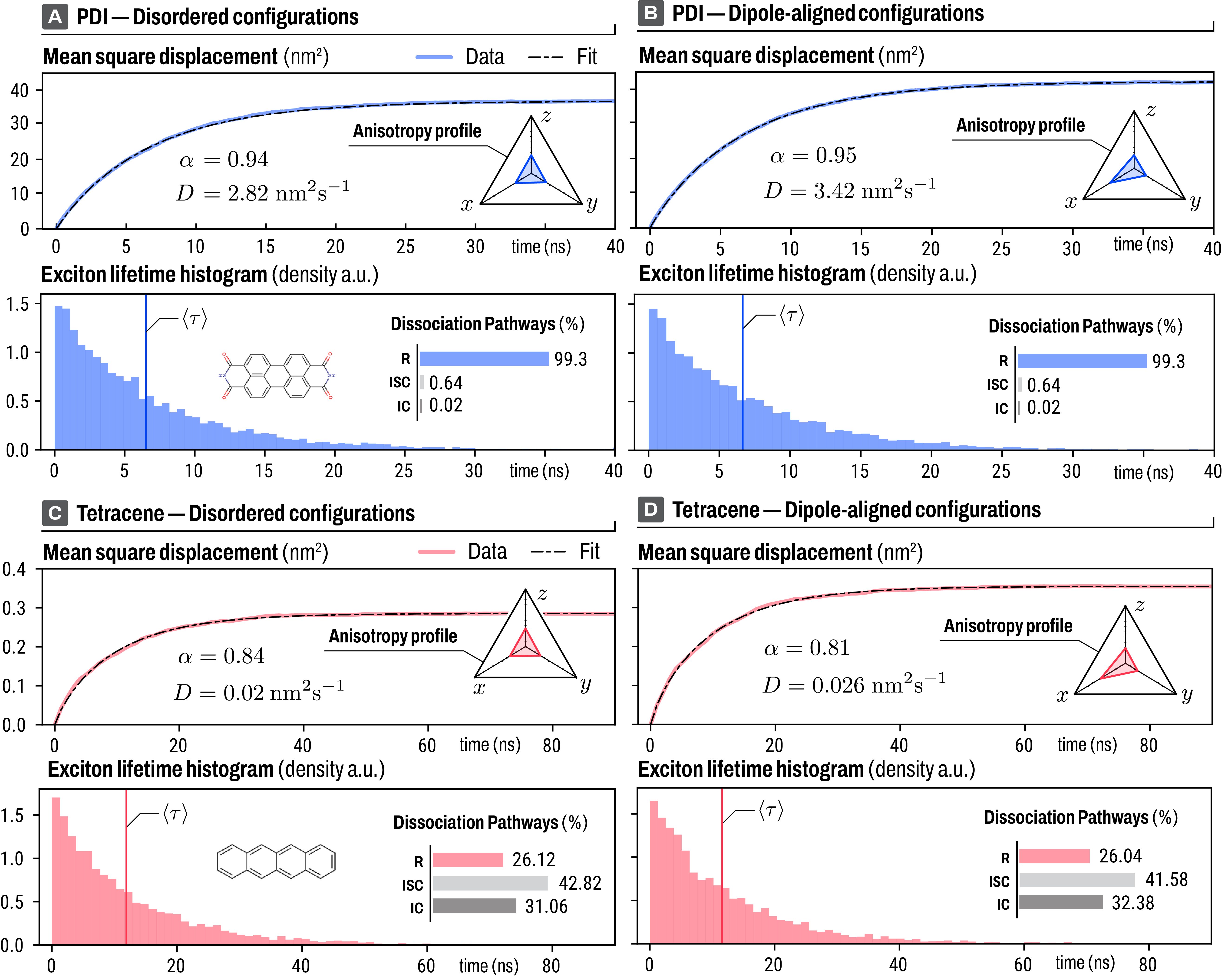}
		\caption{Exciton transport in single-dopant substrates for disordered (\emph{left}) and dipole-aligned (\emph{right}) configuration. Singlet transport is more efficient in PDI (\emph{top}) than in Tetracene (\emph{bottom}), despite the former being characterised by shorter exciton lifetimes. Dipole-aligned configurations have enhanced transport properties and a slight anisotropy towards the dipoles' orientation, which coincides with the $x$-axis. Generalised diffusion coefficient $D$ and diffusion anomaly $\alpha$ are fitted to the data as discussed in Sec.~\ref{ss:exciton_transport_properties}.}
		\label{fig:single-dopant}
		\hrulefill
	\end{figure*}
	
	\begin{figure*}[t]
		\centering	\includegraphics[width=\linewidth]{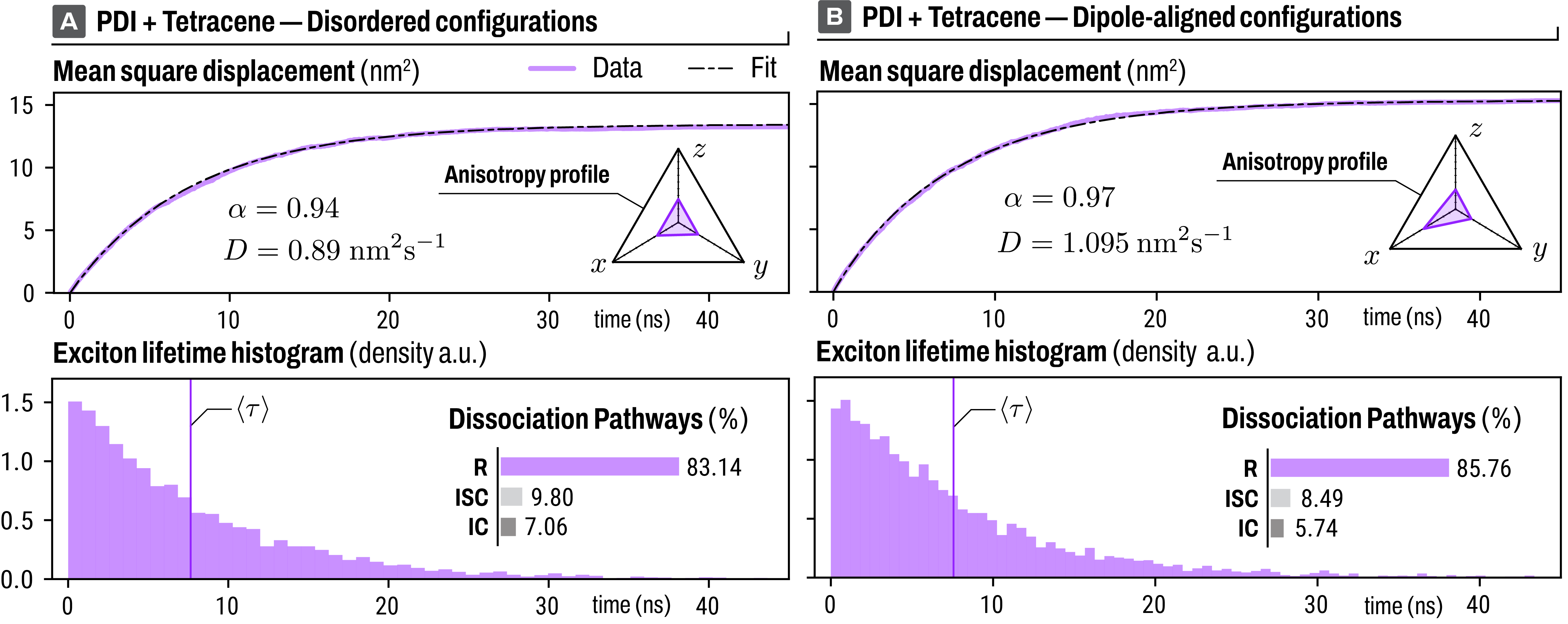}
		\caption{Exciton transport in multi-dopant substrates for disordered (\emph{left}) and dipole-aligned (\emph{right}) configuration. Singlet transport in PDI-Tetracene mixtures is vastly more efficient than in Tetracene, as excitons spend the majority of their lifetime on PDI chromophores. As for the case of single-dopant substrates, dipole-aligned configurations have enhanced transport properties and display anisotropy towards the dipoles' orientation, given by the $x$-axis. Generalised diffusion coefficient $D$ and diffusion anomaly $\alpha$ are fitted to the data as discussed in Sec.~\ref{ss:exciton_transport_properties}.}
		\label{fig:multi-dopant}
		\hrulefill
	\end{figure*}
	
	The results of KMC simulation were analysed to obtain average transport properties. The mean square displacement
	\begin{equation}
		\label{eq:msd}
		\langle r^2(t)\rangle  = \frac{1}{N}\sum_{i=1}^{N}\lVert \vec{r}_i(t) - \vec{r}_i(t_0)\rVert^2, 
	\end{equation}
	was calculated for each ensemble of $N = 5000$ trajectories, where $\vec{r}_i(t)$ is the position of the exciton at time $t$ for the $i$-th trajectory. The mean square displacement was then used to infer the diffusion transport properties according to the following model:
	\begin{equation}
		\label{eq:general_transport}
		\langle r^2(t)\rangle  =2 D \alpha^2 k^{-\alpha} \gamma(\alpha, k t) \frac{\Gamma(\alpha)}{\Gamma(\alpha + 1)},
	\end{equation}
	which represents the time-evolution of the mean square displacement for anomalous diffusion with \textit{generalised} diffusion coefficient $D$, diffusion anomaly $\alpha$ (regular diffusion for $\alpha = 1$, sub-diffusive transport for $\alpha < 1$), and dissociation rate $k$, while the special functions  $\Gamma$ and $\gamma$ in Eq.~\eqref{eq:general_transport} are the gamma and the lower incomplete gamma functions, respectively (see Supplementary Information for the derivation of the model).
	
	Exciton transport in disordered materials is well known to have diffusive to sub-diffusive character~\cite{Akselrod2014}. To accurately infer the anomaly $\alpha$ of the considered diffusion processes we also simulated dissociation-free exciton transport. This was done by removing the dissociation rates from the KMC routine, and sampling 5000 trajectories until $t > 2\tau_\text{diss}$, with $\tau_\text{diss}^{-1} = k_r + k_{IC} + k_{ISC}$ being the theoretical exciton lifetime of the considered dye (or the largest dissociation lifetime for the case of mixed dyes). The mean square displacement obtained from these ensembles was used to fit the anomaly $\alpha$ using the following dissociation-free model
	\begin{equation}
		\label{eq:diffusion_anomaly}
		\langle r^2(t)\rangle  \propto t^\alpha.
	\end{equation}
	
	The model of Eq.~\eqref{eq:general_transport} was then used to fit the remaining parameters $D$ and $k$, by fixing the anomaly $\alpha = \alpha_\texttt{Fit}$ gathered from the dissociation-free trajectories. For single-dopant substrates, the dissociation rate is expected to be equal to $k_r + k_{IC} + k_{ISC}$. This allows us to use Eq.~\eqref{eq:general_transport} as a consistency check of the transport model, since one has to expect $k_\texttt{Fit} \approx k_r + k_{IC} + k_{ISC}$. 
	The mean exciton lifetime $\langle \tau \rangle$ was also calculated from the KMC trajectories,
	\begin{equation}
		\langle \tau \rangle  = \frac{1}{N}\sum_{i=1}^{N} \tau_i,
	\end{equation}
	with $\tau_i$ being the time at which dissociation occurred for the $i$-th trajectory. For single-dopant substrates $\langle \tau \rangle$, $\tau_{\text{diss}}$ and $k^{-1}_\texttt{Fit}$ are all expected to be in good agreement with each other, providing a consistency condition for the transport simulation and model.
	
	In order to gather information on the anisotropy of the transport process, each cartesian component of the square displacement for the dissociation-free trajectories was studied to obtain $D_x$, $D_y$ and $D_z$, where $D = D_x + D_y + D_z$, $\langle r^2(t)\rangle = \langle x^2(t)\rangle + \langle y^2(t)\rangle+ \langle z^2(t)\rangle$ and where
	\begin{equation}
		\langle x^2(t)\rangle = \frac{1}{N}\sum_{i=1}^{N}|x_i(t) - x_i(t_0)|^2, 
	\end{equation}
	and equivalently for $y$ and $z$. The anisotropy profile, shown in Table~\ref{Results_disordered}, was then determined comparing the relative magnitude of each component with respect to their sum, $D_j / D$ for $j=x,y,z$.
	
	Dissociation pathways, sites and dyes were also analysed. For single-dopant substrate the frequency of each dissociation pathway ($r$, $IC$ or $ISC$) is expected to be in agreement with its respective dissociation rate ($k_r$, $k_{IC}$ or $k_{ISC}$). For multi-dopant substrates instead, KMC simulation results provide  insight on the interplay between different chromophore species, valuable to guide the design of efficient LSCs.
	
	\subsubsection{Disordered configuration}
	The exciton transport properties for disordered configurations are presented in Table~\ref{Results_disordered} and Figures~\ref{fig:single-dopant} and~\ref{fig:multi-dopant}. In disordered substrates transport is isotropic, as expected, due to the random relative orientation of chromophores. Single-dopant PDI substrates display the best transport properties, with an average diffusion length $\langle r \rangle = 4.76\; \text{nm}$, near-unit anomaly $\alpha = 0.94$, and diffusion coefficient $D = 2.82\; \text{nm}^2/\text{ns}$, despite having the shortest exciton lifetime $\langle \tau \rangle  = 6.51\; \text{ns}$. As expected from the PLQY, PDI's dissociation is predominately radiative (99.34\%), with marginal contributions from IC (0.65\%) and ISC (0.02\%) recombination pathways; a desirable property for the realisation of LSCs. The mean exciton lifetime $\langle \tau\rangle = 6.51 \; \text{ns} $ and that obtained  by fitting to the model of Eq.~\eqref{eq:general_transport} $\tau_\texttt{fit} = k_\texttt{fit}^{-1} = 6.72\: \text{ns}$ are in good agreement with PDI's theoretical lifetime $\tau_\text{diss} = k_{\text{diss}}^{-1} = 6.60\; \text{ns}$, confirming the validity of the model.
	
	Transport in Tetracene is sub-diffusive, with $\alpha = 0.84$, and diffusion coefficient $D = 0.02\; \text{nm}^2/\text{ns}$. The average diffusion length $\langle r \rangle = 0.46 \;\text{nm}$ is about one order of magnitude smaller than PDI, despite the longer exciton lifetime $\langle \tau \rangle  = 11.89\; \text{ns}$. For comparison, singlet-exciton diffusion in Tetracene crystals is around 12 $\text{nm}$~\cite{Xie2016}. As opposed to PDI, Tetracene dissociation is dominated by IC (42.82\%) and ISC (31.06\%), while radiative recombination only accounts for 26.12\% of the dissociation events, establishing it as a poor choice for single-dopant substrate LSCs. 
	The mean exciton lifetime $\langle \tau\rangle = 11.89 \; \text{ns} $ and that obtained by fitting to the model of Eq.~\eqref{eq:general_transport} $\tau_\texttt{fit} = k_\texttt{fit}^{-1} = 11.62\: \text{ns}$ are in good agreement with Tetracene's theoretical lifetime $\tau_\text{diss} = k_{\text{diss}}^{-1} = 11.74\; \text{ns}$, confirming the validity of the model.
	
	\begin{figure*}[t]
		\centering	\includegraphics[width=\linewidth]{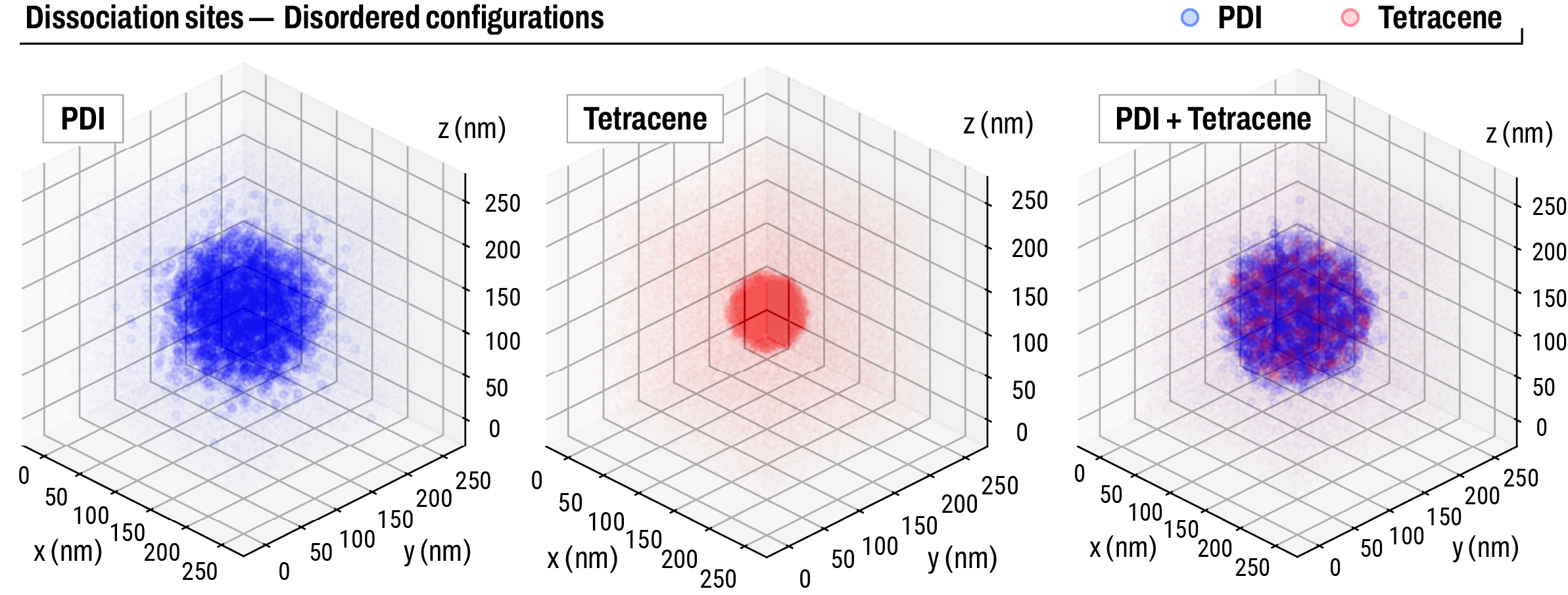}
		\caption{Dissociation sites in single and multi-dopant substrates with disordered configurations for samples of 5000 exciton trajectories. All exciton trajectories start in the centre of the simulation box. Remarkably, exciton transport properties of PDI-Tetracene mixtures are almost as good as in PDI, and vastly better than in Tetracene. As a result,  mixed-dye materials provide a PDI-Tetracene \textit{singlet-bus} mechanism which allows excitons to reach Tetracene sites that would otherwise be out of range in single-dopant substrates (see the Supplementary Information for the case of dipole-aligned configurations).}
		\label{fig:diss_sites1}
		\hrulefill
	\end{figure*}
	
	The transport properties of the multi-dopant PDI+Tetracene substrate are remarkably unbalanced towards PDI's properties. This is due to the fact that the excitons spend the majority of their lifetime ($\langle \tau \rangle  \approx 7.65\; \text{ns}$) on PDI chromophores, as expected from the strong asymmetry in the inter-species FRET rates shown in Table~\ref{TableOverlap}. Transport is near-diffusive with $\alpha = 0.94$, diffusion coefficient $D = 0.89\; \text{nm}^2/\text{ns}$, and  average diffusion length $\langle r \rangle = 2.78 \;\text{nm}$. Radiative dissociation (83.14\%) is remarkably more frequent than  IC (9.82\%) and ISC (7.06\%), returning an effective PLQY of around 0.83. While the PDI+Tetracene transport properties do not seem to favor a multi-dopant substrates over single-dopant ones, they  suggest that mixed-dyes LSCs could prove effective for harnessing singlet excitons, from PDI, and up-converted triplets, from Tetracene, while avoiding Tetracene-induced singlet-quenching. The mean exciton lifetime $\langle \tau\rangle = 7.65 \; \text{ns} $ is in excellent agreement with that obtained by fitting to the model of Eq.~\eqref{eq:general_transport} $\tau_\texttt{fit} = k_\texttt{fit}^{-1} = 7.65\: \text{ns}$, confirming the validity of the model.
	
	\subsubsection{Dipole-aligned configuration}
	
	The exciton transport properties for dipole-aligned configurations are summarised in Table~\ref{Results_disordered}. In  dipole-aligned substrates transport is mildly anisotropic. Perfect anisotropy, characteristic of purely crystalline materials, is not to be expected, as the orientation factor $\kappa$ depends on donor and acceptor dipoles and their orientation with respect to the inter-chromophore separation. Here, instead, the combination of aligned dipoles and randomly distributed sites returns an intermediate regime of anisotropy. For both single-dopant and multi-dopant system the anisotropy is unbalanced towards the $x$ direction, i.e., the typical dipole orientation, with no preferential contribution between the $y$ and $z$ components. 
	
	Exciton diffusion is noticeably enhanced by the alignment of the dipoles, with improved transport properties across all considered substrates. As for the case of disordered configurations, single-dopant PDI substrates display the best transport properties, with an average diffusion length $\langle r \rangle = 5.23\; \text{nm}$, near-unit anomaly $\alpha = 0.95$, and diffusion coefficient $D = 3.43\; \text{nm}^2/\text{ns}$, despite having the shortest exciton lifetime $\langle \tau \rangle  = 6.66\; \text{ns}$. PDI's dissociation pathways are unchanged with respect to the case of disordered configurations. Numerical, fitted and theoretical exciton lifetimes are in good agreement with each other, with $\langle \tau\rangle = 6.66 \; \text{ns}$, $\tau_\texttt{fit} = k_\texttt{fit}^{-1} = 6.69\: \text{ns}$ and $\tau_\text{diss} = k_{\text{diss}}^{-1} = 6.60\; \text{ns}$.
	
	Transport in Tetracene is sub-diffusive, with $\alpha = 0.81$, and diffusion coefficient $D = 0.03\; \text{nm}^2/\text{ns}$. The average diffusion length $\langle r \rangle = 0.5 \;\text{nm}$ is only marginally improved by the alignment of the dipoles, and the dissociation pathways are unaffected. Numerical, fitted and theoretical exciton lifetimes are in good agreement with each other, with $\langle \tau\rangle = 11.89 \; \text{ns}$, $\tau_\texttt{fit} = k_\texttt{fit}^{-1} = 11.88\: \text{ns}$ and $\tau_\text{diss} = k_{\text{diss}}^{-1} = 11.74\; \text{ns}$.
	
	Similar improvement to the transport properties are also obtained for the case of PDI+Tetracene substrates. Transport is near-diffusive with $\alpha = 0.97$, diffusion coefficient $D = 1.10\; \text{nm}^2/\text{ns}$, and average diffusion length $\langle r \rangle = 3.15 \;\text{nm}$. Radiative dissociation (85.76\%) remains dominant over IC (8.49\%) and ISC (5.74\%), returning an effective PLQY of around 0.85. Numerical and fitted exciton lifetimes are in excellent agreement with each other, with $\langle \tau\rangle = 7.56\; \text{ns}$, $\tau_\texttt{fit} = k_\texttt{fit}^{-1} = 7.56\: \text{ns}$.
	
	It should be noted that the direction of photon emission during a fluorescence event is directly related to the direction of transition dipole alignment\cite{tdmoifospfoitioma}. While in the case of a truly disordered system, this propagation is random, in the case of a dipole-aligned system, emitted photons can all be considered to propagate in the same direction. However, aligned-dipole systems have been shown to increase the probability of molecular aggregation\cite{tdmoifospfoitioma}, resulting in an increased chance of excitonic quenching pathways\cite{aceoaieqmfpdi}. As such, some work into the molecular dynamics of dipole-aligned heterogeneous systems may need to be considered, in addition to bulkier derivative which mitigate aggregative tendencies. 
	
	These results highlight the possible advantages of using multi-dopant substrates. Here, we have shown that for heterogeneous PDI-Tetracene substrates, transport is mostly mediated by a particular species of chromophore (PDI). Although transport tends to terminate upon such dye, it is possible for it to also end on Tetracene chromophores. More importantly, it can end on Tetracene chromophores that are much further away from the original excitation site than for single-dopant Tetracene materials, as shown in Figure~\ref{fig:diss_sites1}. As a result, mixed-dye materials provide a PDI-Tetracene \textit{singlet-bus} mechanism which allows excitons to reach Tetracene sites that would otherwise be out of range in single-dopant Tetracene substrates. 
	
	\section{Conclusion}
	
	Quantum chemical analysis revealed very strong emission properties for PDI, in contrast to weaker emission properties for Tetracene. Simulated homogeneous substrates were observed to display strong spectral overlaps, resulting in diffusion properties which agreed well with the literature.  Heterogeneous PDI-Tetracene systems on the other hand, displayed a favouritism between diffusion sites. Here, singlet excitons residing on Tetracene sites were observed to have higher probabilities to hop to PDI sites. 
	
	The KMC simulations provided a fast, computationally inexpensive and reliable approach to simulate the transport properties of singlet excitons in single and multi-dopant substrates. While some exciton transport properties of single-dopant substrates can be predicted analytically, multi-dye materials often necessitate a numerical treatment. The results presented in this work have been entirely obtained from ab-initio calculations and did not require any experimental nor phenomenological parametrisation. This establishes our approach as a powerful technique for LSCs design with existing and yet-to-be synthesised dyes. Furthermore, the approach used in this work can be extended to other exciton pathways, assuming that the corresponding rates are known or can be calculated.
	
	While here we have only considered singlet exciton transport, the inclusion of triplet excitons dynamics highlights another benefit of multi-dopant substrates, i.e., the harvesting of a broader portion of the available spectrum via upconversion pathways. This can be achieved by engineering substrates containing chromophores with efficient singlet transport ($A$), and dyes with efficient triplet transport and up-conversion mechanism ($B$), such that the \textit{singlet-bus} can be utilised to maximize device efficiency. The dyes would be required to favour FRET transport of singlets from $B$ to $A$, and Dexter transport of triplets from $A$ to $B$. In such way, singlet excitons would mostly undergo FRET transport in $A$, while triplet excitons can be recycled into higher-energy singlets instead of being lost to non-radiative pathways. The approach used in this work can be adapted to address the design of these materials, and we expect it to be used in future studies. Our outlook to build on this work will involve adding in ray tracing elements, and multi-exciton interactions, so that we can push towards an even more accurate model. 
	
	\section{Supplementary Information}
	Supplementary information contains the method used to ascertain a chromophore concentration analogous to the low concentration scheme, pictorial representation of the culling radius, and a derivation of the anomalous diffusion with dissociation rates. We also present the dissociation sites for dipole-aligned configurations. Finally, the optimised cartesian coordinates of all relevant electronic states are given. 
	\section{Authors' Contributions}
	A.M. and F.C. contributed equally to this work.
	
	\section{Competing Interests}
	This work was supported by the Australian Government through the Australian Research Council (ARC) under the Centre of Excellence scheme (project number CE170100026). This work was also supported by computational resources provided by the Australian Government through the National Computational Infrastructure National Facility and the Pawsay Supercomputer Centre. 
	
	\section{Data Availability}
	The data that supports the findings of this study are available within the article [and its Supplementary Information]. 
	
	\section{References}
		\bibliography{Paper4v9.bib}

\begin{thebibliography}{74}%
\makeatletter
\providecommand \@ifxundefined [1]{%
 \@ifx{#1\undefined}
}%
\providecommand \@ifnum [1]{%
 \ifnum #1\expandafter \@firstoftwo
 \else \expandafter \@secondoftwo
 \fi
}%
\providecommand \@ifx [1]{%
 \ifx #1\expandafter \@firstoftwo
 \else \expandafter \@secondoftwo
 \fi
}%
\providecommand \natexlab [1]{#1}%
\providecommand \enquote  [1]{``#1''}%
\providecommand \bibnamefont  [1]{#1}%
\providecommand \bibfnamefont [1]{#1}%
\providecommand \citenamefont [1]{#1}%
\providecommand \href@noop [0]{\@secondoftwo}%
\providecommand \href [0]{\begingroup \@sanitize@url \@href}%
\providecommand \@href[1]{\@@startlink{#1}\@@href}%
\providecommand \@@href[1]{\endgroup#1\@@endlink}%
\providecommand \@sanitize@url [0]{\catcode `\\12\catcode `\$12\catcode
  `\&12\catcode `\#12\catcode `\^12\catcode `\_12\catcode `\%12\relax}%
\providecommand \@@startlink[1]{}%
\providecommand \@@endlink[0]{}%
\providecommand \url  [0]{\begingroup\@sanitize@url \@url }%
\providecommand \@url [1]{\endgroup\@href {#1}{\urlprefix }}%
\providecommand \urlprefix  [0]{URL }%
\providecommand \Eprint [0]{\href }%
\providecommand \doibase [0]{https://doi.org/}%
\providecommand \selectlanguage [0]{\@gobble}%
\providecommand \bibinfo  [0]{\@secondoftwo}%
\providecommand \bibfield  [0]{\@secondoftwo}%
\providecommand \translation [1]{[#1]}%
\providecommand \BibitemOpen [0]{}%
\providecommand \bibitemStop [0]{}%
\providecommand \bibitemNoStop [0]{.\EOS\space}%
\providecommand \EOS [0]{\spacefactor3000\relax}%
\providecommand \BibitemShut  [1]{\csname bibitem#1\endcsname}%
\let\auto@bib@innerbib\@empty
\bibitem [{\citenamefont {Banal}\ \emph {et~al.}(2016)\citenamefont {Banal},
  \citenamefont {Zhang}, \citenamefont {Jones}, \citenamefont {Ghiggino},\ and\
  \citenamefont {Wong}}]{emaaemilsc}%
  \BibitemOpen
  \bibfield  {author} {\bibinfo {author} {\bibfnamefont {J.~L.}\ \bibnamefont
  {Banal}}, \bibinfo {author} {\bibfnamefont {B.}~\bibnamefont {Zhang}},
  \bibinfo {author} {\bibfnamefont {D.~J.}\ \bibnamefont {Jones}}, \bibinfo
  {author} {\bibfnamefont {K.~P.}\ \bibnamefont {Ghiggino}},\ and\ \bibinfo
  {author} {\bibfnamefont {W.~W.~H.}\ \bibnamefont {Wong}},\ }\bibfield
  {title} {\enquote {\bibinfo {title} {Emissive molecular aggregates and energy
  migration in luminescent solar concentrators},}\ }\href
  {https://doi.org/10.1021/acs.accounts.6b00432} {\bibfield  {journal}
  {\bibinfo  {journal} {Acc Chem Res}\ }\textbf {\bibinfo {volume} {50}},\
  \bibinfo {pages} {49--57} (\bibinfo {year} {2016})}\BibitemShut {NoStop}%
\bibitem [{\citenamefont {Debije}\ and\ \citenamefont
  {Verbunt}(2011)}]{tyolscr}%
  \BibitemOpen
  \bibfield  {author} {\bibinfo {author} {\bibfnamefont {M.~G.}\ \bibnamefont
  {Debije}}\ and\ \bibinfo {author} {\bibfnamefont {P.~P.~C.}\ \bibnamefont
  {Verbunt}},\ }\bibfield  {title} {\enquote {\bibinfo {title} {Thirty years of
  luminescent solar concentrator research: Solar energy for the built
  environment},}\ }\href {https://doi.org/10.1002/aenm.201100554} {\bibfield
  {journal} {\bibinfo  {journal} {Advanced Energy Materials}\ }\textbf
  {\bibinfo {volume} {2}},\ \bibinfo {pages} {12--35} (\bibinfo {year}
  {2011})}\BibitemShut {NoStop}%
\bibitem [{\citenamefont {McKenna}\ and\ \citenamefont
  {Evans}(2017)}]{tesctmdflsd}%
  \BibitemOpen
  \bibfield  {author} {\bibinfo {author} {\bibfnamefont {B.}~\bibnamefont
  {McKenna}}\ and\ \bibinfo {author} {\bibfnamefont {R.~C.}\ \bibnamefont
  {Evans}},\ }\bibfield  {title} {\enquote {\bibinfo {title} {Towards efficient
  spectral converters through materials design for luminescent solar
  devices},}\ }\href {https://doi.org/10.1002/adma.201606491} {\bibfield
  {journal} {\bibinfo  {journal} {Advanced Materials}\ }\textbf {\bibinfo
  {volume} {29}},\ \bibinfo {pages} {1606491} (\bibinfo {year}
  {2017})}\BibitemShut {NoStop}%
\bibitem [{\citenamefont {Zhang}\ \emph
  {et~al.}(2020{\natexlab{a}})\citenamefont {Zhang}, \citenamefont {Yang},
  \citenamefont {Warner}, \citenamefont {Mulvaney}, \citenamefont
  {Rosengarten}, \citenamefont {Wong},\ and\ \citenamefont
  {Ghiggino}}]{alscrtswagui}%
  \BibitemOpen
  \bibfield  {author} {\bibinfo {author} {\bibfnamefont {B.}~\bibnamefont
  {Zhang}}, \bibinfo {author} {\bibfnamefont {H.}~\bibnamefont {Yang}},
  \bibinfo {author} {\bibfnamefont {T.}~\bibnamefont {Warner}}, \bibinfo
  {author} {\bibfnamefont {P.}~\bibnamefont {Mulvaney}}, \bibinfo {author}
  {\bibfnamefont {G.}~\bibnamefont {Rosengarten}}, \bibinfo {author}
  {\bibfnamefont {W.~W.~H.}\ \bibnamefont {Wong}},\ and\ \bibinfo {author}
  {\bibfnamefont {K.~P.}\ \bibnamefont {Ghiggino}},\ }\bibfield  {title}
  {\enquote {\bibinfo {title} {A luminescent solar concentrator ray tracing
  simulator with a graphical user interface: features and applications},}\
  }\href {https://doi.org/10.1088/2050-6120/ab993d} {\bibfield  {journal}
  {\bibinfo  {journal} {Methods and Applications in Fluorescence}\ }\textbf
  {\bibinfo {volume} {8}},\ \bibinfo {pages} {037001} (\bibinfo {year}
  {2020}{\natexlab{a}})}\BibitemShut {NoStop}%
\bibitem [{\citenamefont {van Eersel}, \citenamefont {Bobbert},\ and\
  \citenamefont {Coehoorn}(2015)}]{Eersel2015}%
  \BibitemOpen
  \bibfield  {author} {\bibinfo {author} {\bibfnamefont {H.}~\bibnamefont {van
  Eersel}}, \bibinfo {author} {\bibfnamefont {P.~A.}\ \bibnamefont {Bobbert}},\
  and\ \bibinfo {author} {\bibfnamefont {R.}~\bibnamefont {Coehoorn}},\
  }\bibfield  {title} {\enquote {\bibinfo {title} {{Kinetic Monte Carlo study
  of triplet-triplet annihilation in organic phosphorescent emitters}},}\
  }\href {https://doi.org/10.1063/1.4914460} {\bibfield  {journal} {\bibinfo
  {journal} {Journal of Applied Physics}\ }\textbf {\bibinfo {volume} {117}},\
  \bibinfo {pages} {115502} (\bibinfo {year} {2015})}\BibitemShut {NoStop}%
\bibitem [{\citenamefont {Mou}\ \emph {et~al.}(2013)\citenamefont {Mou},
  \citenamefont {Hattori}, \citenamefont {Rajak}, \citenamefont {Shimojo},\
  and\ \citenamefont {Nakano}}]{nmosfiams}%
  \BibitemOpen
  \bibfield  {author} {\bibinfo {author} {\bibfnamefont {W.}~\bibnamefont
  {Mou}}, \bibinfo {author} {\bibfnamefont {S.}~\bibnamefont {Hattori}},
  \bibinfo {author} {\bibfnamefont {P.}~\bibnamefont {Rajak}}, \bibinfo
  {author} {\bibfnamefont {F.}~\bibnamefont {Shimojo}},\ and\ \bibinfo {author}
  {\bibfnamefont {A.}~\bibnamefont {Nakano}},\ }\bibfield  {title} {\enquote
  {\bibinfo {title} {Nanoscopic mechanisms of singlet fission in amorphous
  molecular solid},}\ }\href {https://doi.org/10.1063/1.4795138} {\bibfield
  {journal} {\bibinfo  {journal} {Applied Physics Letters}\ }\textbf {\bibinfo
  {volume} {102}},\ \bibinfo {pages} {173301} (\bibinfo {year}
  {2013})}\BibitemShut {NoStop}%
\bibitem [{\citenamefont {Nakano}\ \emph {et~al.}(2019)\citenamefont {Nakano},
  \citenamefont {Okada}, \citenamefont {Nagami}, \citenamefont {Tonami},
  \citenamefont {Kishi},\ and\ \citenamefont {Kitagawa}}]{mcwatkfdoma}%
  \BibitemOpen
  \bibfield  {author} {\bibinfo {author} {\bibfnamefont {M.}~\bibnamefont
  {Nakano}}, \bibinfo {author} {\bibfnamefont {K.}~\bibnamefont {Okada}},
  \bibinfo {author} {\bibfnamefont {T.}~\bibnamefont {Nagami}}, \bibinfo
  {author} {\bibfnamefont {T.}~\bibnamefont {Tonami}}, \bibinfo {author}
  {\bibfnamefont {R.}~\bibnamefont {Kishi}},\ and\ \bibinfo {author}
  {\bibfnamefont {Y.}~\bibnamefont {Kitagawa}},\ }\bibfield  {title} {\enquote
  {\bibinfo {title} {Monte carlo wavefunction approach to singlet fission
  dynamics of molecular aggregates},}\ }\href
  {https://doi.org/10.3390/molecules24030541} {\bibfield  {journal} {\bibinfo
  {journal} {Molecules}\ }\textbf {\bibinfo {volume} {24}},\ \bibinfo {pages}
  {541} (\bibinfo {year} {2019})}\BibitemShut {NoStop}%
\bibitem [{\citenamefont {Ansari-Rad}(2021)}]{taaottiopsmkcmam}%
  \BibitemOpen
  \bibfield  {author} {\bibinfo {author} {\bibfnamefont {M.}~\bibnamefont
  {Ansari-Rad}},\ }\bibfield  {title} {\enquote {\bibinfo {title} {Transport
  and annihilation of the triplets in organic phosphorescent systems: Kinetic
  monte carlo simulation and modeling},}\ }\href
  {https://doi.org/10.1021/acs.jpcc.0c10138} {\bibfield  {journal} {\bibinfo
  {journal} {The Journal of Physical Chemistry C}\ }\textbf {\bibinfo {volume}
  {125}},\ \bibinfo {pages} {5760--5770} (\bibinfo {year} {2021})}\BibitemShut
  {NoStop}%
\bibitem [{\citenamefont {Xie}\ and\ \citenamefont
  {Ma}(2016{\natexlab{a}})}]{oaeosateditc}%
  \BibitemOpen
  \bibfield  {author} {\bibinfo {author} {\bibfnamefont {X.}~\bibnamefont
  {Xie}}\ and\ \bibinfo {author} {\bibfnamefont {H.}~\bibnamefont {Ma}},\
  }\bibfield  {title} {\enquote {\bibinfo {title} {Opposite anisotropy effects
  of singlet and triplet exciton diffusion in tetracene crystal},}\ }\href
  {https://doi.org/10.1002/open.201500214} {\bibfield  {journal} {\bibinfo
  {journal} {{ChemistryOpen}}\ }\textbf {\bibinfo {volume} {5}},\ \bibinfo
  {pages} {201--205} (\bibinfo {year} {2016}{\natexlab{a}})}\BibitemShut
  {NoStop}%
\bibitem [{\citenamefont {Zhang}\ \emph
  {et~al.}(2017{\natexlab{a}})\citenamefont {Zhang}, \citenamefont
  {Soleimaninejad}, \citenamefont {Jones}, \citenamefont {White}, \citenamefont
  {Ghiggino}, \citenamefont {Smith},\ and\ \citenamefont
  {Wong}}]{hfmipdeocopp}%
  \BibitemOpen
  \bibfield  {author} {\bibinfo {author} {\bibfnamefont {B.}~\bibnamefont
  {Zhang}}, \bibinfo {author} {\bibfnamefont {H.}~\bibnamefont
  {Soleimaninejad}}, \bibinfo {author} {\bibfnamefont {D.~J.}\ \bibnamefont
  {Jones}}, \bibinfo {author} {\bibfnamefont {J.~M.}\ \bibnamefont {White}},
  \bibinfo {author} {\bibfnamefont {K.~P.}\ \bibnamefont {Ghiggino}}, \bibinfo
  {author} {\bibfnamefont {T.~A.}\ \bibnamefont {Smith}},\ and\ \bibinfo
  {author} {\bibfnamefont {W.~W.~H.}\ \bibnamefont {Wong}},\ }\bibfield
  {title} {\enquote {\bibinfo {title} {Highly fluorescent molecularly insulated
  perylene diimides: Effect of concentration on photophysical properties},}\
  }\href {https://doi.org/10.1021/acs.chemmater.7b02968} {\bibfield  {journal}
  {\bibinfo  {journal} {Chem Mater}\ }\textbf {\bibinfo {volume} {29}},\
  \bibinfo {pages} {8395--8403} (\bibinfo {year}
  {2017}{\natexlab{a}})}\BibitemShut {NoStop}%
\bibitem [{\citenamefont {Tummeltshammer}\ \emph {et~al.}(2016)\citenamefont
  {Tummeltshammer}, \citenamefont {Taylor}, \citenamefont {Kenyon},\ and\
  \citenamefont {Papakonstantinou}}]{lilscu}%
  \BibitemOpen
  \bibfield  {author} {\bibinfo {author} {\bibfnamefont {C.}~\bibnamefont
  {Tummeltshammer}}, \bibinfo {author} {\bibfnamefont {A.}~\bibnamefont
  {Taylor}}, \bibinfo {author} {\bibfnamefont {A.}~\bibnamefont {Kenyon}},\
  and\ \bibinfo {author} {\bibfnamefont {I.}~\bibnamefont {Papakonstantinou}},\
  }\bibfield  {title} {\enquote {\bibinfo {title} {Losses in luminescent solar
  concentrators unveiled},}\ }\href
  {https://doi.org/10.1016/j.solmat.2015.08.008} {\bibfield  {journal}
  {\bibinfo  {journal} {Sol Energ Mat Sol}\ }\textbf {\bibinfo {volume}
  {144}},\ \bibinfo {pages} {40--47} (\bibinfo {year} {2016})}\BibitemShut
  {NoStop}%
\bibitem [{\citenamefont {Zhang}\ \emph
  {et~al.}(2020{\natexlab{b}})\citenamefont {Zhang}, \citenamefont {Lyskov},
  \citenamefont {Wilson}, \citenamefont {Sabatini}, \citenamefont {Manian},
  \citenamefont {Soleimaninejad}, \citenamefont {White}, \citenamefont {Smith},
  \citenamefont {Lakhwani}, \citenamefont {Jones}, \citenamefont {Ghiggino},
  \citenamefont {Russo},\ and\ \citenamefont {Wong}}]{FRETeploPDIbcmaai}%
  \BibitemOpen
  \bibfield  {author} {\bibinfo {author} {\bibfnamefont {B.}~\bibnamefont
  {Zhang}}, \bibinfo {author} {\bibfnamefont {I.}~\bibnamefont {Lyskov}},
  \bibinfo {author} {\bibfnamefont {L.~J.}\ \bibnamefont {Wilson}}, \bibinfo
  {author} {\bibfnamefont {R.~P.}\ \bibnamefont {Sabatini}}, \bibinfo {author}
  {\bibfnamefont {A.}~\bibnamefont {Manian}}, \bibinfo {author} {\bibfnamefont
  {H.}~\bibnamefont {Soleimaninejad}}, \bibinfo {author} {\bibfnamefont
  {J.~M.}\ \bibnamefont {White}}, \bibinfo {author} {\bibfnamefont {T.~A.}\
  \bibnamefont {Smith}}, \bibinfo {author} {\bibfnamefont {G.}~\bibnamefont
  {Lakhwani}}, \bibinfo {author} {\bibfnamefont {D.~J.}\ \bibnamefont {Jones}},
  \bibinfo {author} {\bibfnamefont {K.~P.}\ \bibnamefont {Ghiggino}}, \bibinfo
  {author} {\bibfnamefont {S.~P.}\ \bibnamefont {Russo}},\ and\ \bibinfo
  {author} {\bibfnamefont {W.~W.~H.}\ \bibnamefont {Wong}},\ }\bibfield
  {title} {\enquote {\bibinfo {title} {{FRET}-enhanced photoluminescence of
  perylene diimides by combining molecular aggregation and insulation},}\
  }\href {https://doi.org/10.1039/d0tc02108c} {\bibfield  {journal} {\bibinfo
  {journal} {J Mater Chem C}\ }\textbf {\bibinfo {volume} {8}},\ \bibinfo
  {pages} {8953--8961} (\bibinfo {year} {2020}{\natexlab{b}})}\BibitemShut
  {NoStop}%
\bibitem [{\citenamefont {Meftahi}\ \emph {et~al.}(2020)\citenamefont
  {Meftahi}, \citenamefont {Manian}, \citenamefont {Christofferson},
  \citenamefont {Lyskov},\ and\ \citenamefont {Russo}}]{aceoaieqmfpdi}%
  \BibitemOpen
  \bibfield  {author} {\bibinfo {author} {\bibfnamefont {N.}~\bibnamefont
  {Meftahi}}, \bibinfo {author} {\bibfnamefont {A.}~\bibnamefont {Manian}},
  \bibinfo {author} {\bibfnamefont {A.~J.}\ \bibnamefont {Christofferson}},
  \bibinfo {author} {\bibfnamefont {I.}~\bibnamefont {Lyskov}},\ and\ \bibinfo
  {author} {\bibfnamefont {S.~P.}\ \bibnamefont {Russo}},\ }\bibfield  {title}
  {\enquote {\bibinfo {title} {A computational exploration of
  aggregation-induced excitonic quenching mechanisms for perylene diimide
  chromophores},}\ }\href {https://doi.org/10.1063/5.0013634} {\bibfield
  {journal} {\bibinfo  {journal} {J. Chem. Phys.}\ }\textbf {\bibinfo {volume}
  {153}},\ \bibinfo {pages} {064108} (\bibinfo {year} {2020})}\BibitemShut
  {NoStop}%
\bibitem [{\citenamefont {Manian}\ \emph {et~al.}(2021)\citenamefont {Manian},
  \citenamefont {Shaw}, \citenamefont {Lyskov}, \citenamefont {Wong},\ and\
  \citenamefont {Russo}}]{NRpaper}%
  \BibitemOpen
  \bibfield  {author} {\bibinfo {author} {\bibfnamefont {A.}~\bibnamefont
  {Manian}}, \bibinfo {author} {\bibfnamefont {R.~A.}\ \bibnamefont {Shaw}},
  \bibinfo {author} {\bibfnamefont {I.}~\bibnamefont {Lyskov}}, \bibinfo
  {author} {\bibfnamefont {W.}~\bibnamefont {Wong}},\ and\ \bibinfo {author}
  {\bibfnamefont {S.~P.}\ \bibnamefont {Russo}},\ }\bibfield  {title} {\enquote
  {\bibinfo {title} {Modeling radiative and non-radiative pathways at both the
  franck{\textendash}condon and herzberg{\textendash}teller approximation
  level},}\ }\href {https://doi.org/10.1063/5.0058643} {\bibfield  {journal}
  {\bibinfo  {journal} {The Journal of Chemical Physics}\ }\textbf {\bibinfo
  {volume} {155}},\ \bibinfo {pages} {054108} (\bibinfo {year}
  {2021})}\BibitemShut {NoStop}%
\bibitem [{\citenamefont {Leow}\ \emph {et~al.}(2013)\citenamefont {Leow},
  \citenamefont {Corrado}, \citenamefont {Osborn},\ and\ \citenamefont
  {Carter}}]{mcrtsolscfbip}%
  \BibitemOpen
  \bibfield  {author} {\bibinfo {author} {\bibfnamefont {S.~W.}\ \bibnamefont
  {Leow}}, \bibinfo {author} {\bibfnamefont {C.}~\bibnamefont {Corrado}},
  \bibinfo {author} {\bibfnamefont {M.}~\bibnamefont {Osborn}},\ and\ \bibinfo
  {author} {\bibfnamefont {S.~A.}\ \bibnamefont {Carter}},\ }\bibfield  {title}
  {\enquote {\bibinfo {title} {Monte carlo ray-tracing simulations of
  luminescent solar concentrators for building integrated photovoltaics},}\
  }in\ \href {https://doi.org/10.1117/12.2024676} {\emph {\bibinfo {booktitle}
  {High and Low Concentrator Systems for Solar Electric Applications
  {VIII}}}},\ \bibinfo {editor} {edited by\ \bibinfo {editor} {\bibfnamefont
  {A.~P.}\ \bibnamefont {Plesniak}}}\ (\bibinfo  {publisher} {{SPIE}},\
  \bibinfo {year} {2013})\BibitemShut {NoStop}%
\bibitem [{\citenamefont {Zhang}\ \emph
  {et~al.}(2017{\natexlab{b}})\citenamefont {Zhang}, \citenamefont {Zhang},
  \citenamefont {Zhang}, \citenamefont {Yan}, \citenamefont {Song},
  \citenamefont {Jun},\ and\ \citenamefont {Chen}}]{tsaaolsBMPLSCb3dmcrtm}%
  \BibitemOpen
  \bibfield  {author} {\bibinfo {author} {\bibfnamefont {F.}~\bibnamefont
  {Zhang}}, \bibinfo {author} {\bibfnamefont {N.-N.}\ \bibnamefont {Zhang}},
  \bibinfo {author} {\bibfnamefont {Y.}~\bibnamefont {Zhang}}, \bibinfo
  {author} {\bibfnamefont {S.}~\bibnamefont {Yan}}, \bibinfo {author}
  {\bibfnamefont {S.}~\bibnamefont {Song}}, \bibinfo {author} {\bibfnamefont
  {B.}~\bibnamefont {Jun}},\ and\ \bibinfo {author} {\bibfnamefont
  {G.}~\bibnamefont {Chen}},\ }\bibfield  {title} {\enquote {\bibinfo {title}
  {Theoretical simulation and analysis of large size {BMP}-{LSC} by 3d monte
  carlo ray tracing model},}\ }\href
  {https://doi.org/10.1088/1674-1056/26/5/054201} {\bibfield  {journal}
  {\bibinfo  {journal} {Chinese Physics B}\ }\textbf {\bibinfo {volume} {26}},\
  \bibinfo {pages} {054201} (\bibinfo {year} {2017}{\natexlab{b}})}\BibitemShut
  {NoStop}%
\bibitem [{\citenamefont {Athanasopoulos}\ \emph {et~al.}(2007)\citenamefont
  {Athanasopoulos}, \citenamefont {Kirkpatrick}, \citenamefont
  {Mart{\'{i}}nez}, \citenamefont {Frost}, \citenamefont {Foden}, \citenamefont
  {Walker},\ and\ \citenamefont {Nelson}}]{Athanasopoulos2007}%
  \BibitemOpen
  \bibfield  {author} {\bibinfo {author} {\bibfnamefont {S.}~\bibnamefont
  {Athanasopoulos}}, \bibinfo {author} {\bibfnamefont {J.}~\bibnamefont
  {Kirkpatrick}}, \bibinfo {author} {\bibfnamefont {D.}~\bibnamefont
  {Mart{\'{i}}nez}}, \bibinfo {author} {\bibfnamefont {J.~M.}\ \bibnamefont
  {Frost}}, \bibinfo {author} {\bibfnamefont {C.~M.}\ \bibnamefont {Foden}},
  \bibinfo {author} {\bibfnamefont {A.~B.}\ \bibnamefont {Walker}},\ and\
  \bibinfo {author} {\bibfnamefont {J.}~\bibnamefont {Nelson}},\ }\bibfield
  {title} {\enquote {\bibinfo {title} {{Predictive study of charge transport in
  disordered semiconducting polymers}},}\ }\href
  {https://doi.org/10.1021/nl0708718} {\bibfield  {journal} {\bibinfo
  {journal} {Nano Letters}\ }\textbf {\bibinfo {volume} {7}},\ \bibinfo {pages}
  {1785--1788} (\bibinfo {year} {2007})}\BibitemShut {NoStop}%
\bibitem [{\citenamefont {Kimber}\ \emph {et~al.}(2012)\citenamefont {Kimber},
  \citenamefont {Wright}, \citenamefont {O'Kane}, \citenamefont {Walker},\ and\
  \citenamefont {Blakesley}}]{Kimber2012}%
  \BibitemOpen
  \bibfield  {author} {\bibinfo {author} {\bibfnamefont {R.~G.~E.}\
  \bibnamefont {Kimber}}, \bibinfo {author} {\bibfnamefont {E.~N.}\
  \bibnamefont {Wright}}, \bibinfo {author} {\bibfnamefont {S.~E.~J.}\
  \bibnamefont {O'Kane}}, \bibinfo {author} {\bibfnamefont {A.~B.}\
  \bibnamefont {Walker}},\ and\ \bibinfo {author} {\bibfnamefont {J.~C.}\
  \bibnamefont {Blakesley}},\ }\bibfield  {title} {\enquote {\bibinfo {title}
  {{Mesoscopic kinetic Monte Carlo modeling of organic photovoltaic device
  characteristics}},}\ }\href {https://doi.org/10.1103/PhysRevB.86.235206}
  {\bibfield  {journal} {\bibinfo  {journal} {Physical Review B}\ }\textbf
  {\bibinfo {volume} {86}},\ \bibinfo {pages} {235206} (\bibinfo {year}
  {2012})}\BibitemShut {NoStop}%
\bibitem [{\citenamefont {Yost}\ \emph {et~al.}(2012)\citenamefont {Yost},
  \citenamefont {Hontz}, \citenamefont {Yeganeh},\ and\ \citenamefont {{Van
  Voorhis}}}]{Yost2012}%
  \BibitemOpen
  \bibfield  {author} {\bibinfo {author} {\bibfnamefont {S.~R.}\ \bibnamefont
  {Yost}}, \bibinfo {author} {\bibfnamefont {E.}~\bibnamefont {Hontz}},
  \bibinfo {author} {\bibfnamefont {S.}~\bibnamefont {Yeganeh}},\ and\ \bibinfo
  {author} {\bibfnamefont {T.}~\bibnamefont {{Van Voorhis}}},\ }\bibfield
  {title} {\enquote {\bibinfo {title} {{Triplet vs singlet energy transfer in
  organic semiconductors: The tortoise and the hare}},}\ }\href
  {https://doi.org/10.1021/jp304433t} {\bibfield  {journal} {\bibinfo
  {journal} {Journal of Physical Chemistry C}\ }\textbf {\bibinfo {volume}
  {116}},\ \bibinfo {pages} {17369--17377} (\bibinfo {year}
  {2012})}\BibitemShut {NoStop}%
\bibitem [{\citenamefont {Akselrod}\ \emph
  {et~al.}(2014{\natexlab{a}})\citenamefont {Akselrod}, \citenamefont {Prins},
  \citenamefont {Poulikakos}, \citenamefont {Lee}, \citenamefont {Weidman},
  \citenamefont {Mork}, \citenamefont {Willard}, \citenamefont
  {Bulovi{\'{c}}},\ and\ \citenamefont {Tisdale}}]{Akselrod2014a}%
  \BibitemOpen
  \bibfield  {author} {\bibinfo {author} {\bibfnamefont {G.~M.}\ \bibnamefont
  {Akselrod}}, \bibinfo {author} {\bibfnamefont {F.}~\bibnamefont {Prins}},
  \bibinfo {author} {\bibfnamefont {L.~V.}\ \bibnamefont {Poulikakos}},
  \bibinfo {author} {\bibfnamefont {E.~M.}\ \bibnamefont {Lee}}, \bibinfo
  {author} {\bibfnamefont {M.~C.}\ \bibnamefont {Weidman}}, \bibinfo {author}
  {\bibfnamefont {A.~J.}\ \bibnamefont {Mork}}, \bibinfo {author}
  {\bibfnamefont {A.~P.}\ \bibnamefont {Willard}}, \bibinfo {author}
  {\bibfnamefont {V.}~\bibnamefont {Bulovi{\'{c}}}},\ and\ \bibinfo {author}
  {\bibfnamefont {W.~A.}\ \bibnamefont {Tisdale}},\ }\bibfield  {title}
  {\enquote {\bibinfo {title} {{Subdiffusive exciton transport in quantum dot
  solids}},}\ }\href {https://doi.org/10.1021/nl501190s} {\bibfield  {journal}
  {\bibinfo  {journal} {Nano Letters}\ }\textbf {\bibinfo {volume} {14}},\
  \bibinfo {pages} {3556--3562} (\bibinfo {year}
  {2014}{\natexlab{a}})}\BibitemShut {NoStop}%
\bibitem [{\citenamefont {Xie}\ and\ \citenamefont
  {Ma}(2016{\natexlab{b}})}]{Xie2016}%
  \BibitemOpen
  \bibfield  {author} {\bibinfo {author} {\bibfnamefont {X.}~\bibnamefont
  {Xie}}\ and\ \bibinfo {author} {\bibfnamefont {H.}~\bibnamefont {Ma}},\
  }\bibfield  {title} {\enquote {\bibinfo {title} {{Opposite Anisotropy Effects
  of Singlet and Triplet Exciton Diffusion in Tetracene Crystal}},}\ }\href
  {https://doi.org/10.1002/open.201500214} {\bibfield  {journal} {\bibinfo
  {journal} {ChemistryOpen}\ }\textbf {\bibinfo {volume} {5}},\ \bibinfo
  {pages} {201--205} (\bibinfo {year} {2016}{\natexlab{b}})}\BibitemShut
  {NoStop}%
\bibitem [{\citenamefont {Kaiser}\ \emph {et~al.}(2018)\citenamefont {Kaiser},
  \citenamefont {Popp}, \citenamefont {Rinderle}, \citenamefont {Albes},\ and\
  \citenamefont {Gagliardi}}]{Kaiser2018}%
  \BibitemOpen
  \bibfield  {author} {\bibinfo {author} {\bibfnamefont {W.}~\bibnamefont
  {Kaiser}}, \bibinfo {author} {\bibfnamefont {J.}~\bibnamefont {Popp}},
  \bibinfo {author} {\bibfnamefont {M.}~\bibnamefont {Rinderle}}, \bibinfo
  {author} {\bibfnamefont {T.}~\bibnamefont {Albes}},\ and\ \bibinfo {author}
  {\bibfnamefont {A.}~\bibnamefont {Gagliardi}},\ }\bibfield  {title} {\enquote
  {\bibinfo {title} {{Generalized Kinetic Monte Carlo Framework for Organic
  Electronics}},}\ }\href {https://doi.org/10.3390/a11040037} {\bibfield
  {journal} {\bibinfo  {journal} {Algorithms}\ }\textbf {\bibinfo {volume}
  {11}},\ \bibinfo {pages} {37} (\bibinfo {year} {2018})}\BibitemShut {NoStop}%
\bibitem [{\citenamefont {Ligthart}\ \emph {et~al.}(2018)\citenamefont
  {Ligthart}, \citenamefont {de~Vries}, \citenamefont {Zhang}, \citenamefont
  {Pols}, \citenamefont {Bobbert}, \citenamefont {van Eersel},\ and\
  \citenamefont {Coehoorn}}]{Ligthart2018}%
  \BibitemOpen
  \bibfield  {author} {\bibinfo {author} {\bibfnamefont {A.}~\bibnamefont
  {Ligthart}}, \bibinfo {author} {\bibfnamefont {X.}~\bibnamefont {de~Vries}},
  \bibinfo {author} {\bibfnamefont {L.}~\bibnamefont {Zhang}}, \bibinfo
  {author} {\bibfnamefont {M.~C. W.~M.}\ \bibnamefont {Pols}}, \bibinfo
  {author} {\bibfnamefont {P.~A.}\ \bibnamefont {Bobbert}}, \bibinfo {author}
  {\bibfnamefont {H.}~\bibnamefont {van Eersel}},\ and\ \bibinfo {author}
  {\bibfnamefont {R.}~\bibnamefont {Coehoorn}},\ }\bibfield  {title} {\enquote
  {\bibinfo {title} {{Effect of Triplet Confinement on Triplet–Triplet
  Annihilation in Organic Phosphorescent Host–Guest Systems}},}\ }\href
  {https://doi.org/10.1002/ADFM.201804618} {\bibfield  {journal} {\bibinfo
  {journal} {Advanced Functional Materials}\ }\textbf {\bibinfo {volume}
  {28}},\ \bibinfo {pages} {1804618} (\bibinfo {year} {2018})}\BibitemShut
  {NoStop}%
\bibitem [{\citenamefont {Saxena}\ \emph {et~al.}(2020)\citenamefont {Saxena},
  \citenamefont {Meier}, \citenamefont {Athanasopoulos}, \citenamefont
  {B{\"{a}}ssler},\ and\ \citenamefont {K{\"{o}}hler}}]{Saxena2020}%
  \BibitemOpen
  \bibfield  {author} {\bibinfo {author} {\bibfnamefont {R.}~\bibnamefont
  {Saxena}}, \bibinfo {author} {\bibfnamefont {T.}~\bibnamefont {Meier}},
  \bibinfo {author} {\bibfnamefont {S.}~\bibnamefont {Athanasopoulos}},
  \bibinfo {author} {\bibfnamefont {H.}~\bibnamefont {B{\"{a}}ssler}},\ and\
  \bibinfo {author} {\bibfnamefont {A.}~\bibnamefont {K{\"{o}}hler}},\
  }\bibfield  {title} {\enquote {\bibinfo {title} {{Kinetic Monte Carlo Study
  of Triplet-Triplet Annihilation in Conjugated Luminescent Materials}},}\
  }\href {https://doi.org/10.1103/PhysRevApplied.14.034050} {\bibfield
  {journal} {\bibinfo  {journal} {Physical Review Applied}\ }\textbf {\bibinfo
  {volume} {14}},\ \bibinfo {pages} {034050} (\bibinfo {year}
  {2020})}\BibitemShut {NoStop}%
\bibitem [{\citenamefont {Marcus}(1956)}]{marcus}%
  \BibitemOpen
  \bibfield  {author} {\bibinfo {author} {\bibfnamefont {R.~A.}\ \bibnamefont
  {Marcus}},\ }\bibfield  {title} {\enquote {\bibinfo {title} {On the theory of
  oxidation-reduction reactions involving electron transfer. i},}\ }\href
  {https://doi.org/10.1063/1.1742723} {\bibfield  {journal} {\bibinfo
  {journal} {J Chem Phys}\ }\textbf {\bibinfo {volume} {24}},\ \bibinfo {pages}
  {966--978} (\bibinfo {year} {1956})}\BibitemShut {NoStop}%
\bibitem [{\citenamefont {Stehr}\ \emph {et~al.}(2011)\citenamefont {Stehr},
  \citenamefont {Pfister}, \citenamefont {Fink}, \citenamefont {Engels},\ and\
  \citenamefont {Deibel}}]{fpcoaccmiosc}%
  \BibitemOpen
  \bibfield  {author} {\bibinfo {author} {\bibfnamefont {V.}~\bibnamefont
  {Stehr}}, \bibinfo {author} {\bibfnamefont {J.}~\bibnamefont {Pfister}},
  \bibinfo {author} {\bibfnamefont {R.~F.}\ \bibnamefont {Fink}}, \bibinfo
  {author} {\bibfnamefont {B.}~\bibnamefont {Engels}},\ and\ \bibinfo {author}
  {\bibfnamefont {C.}~\bibnamefont {Deibel}},\ }\bibfield  {title} {\enquote
  {\bibinfo {title} {First-principles calculations of anisotropic
  charge-carrier mobilities in organic semiconductor crystals},}\ }\href
  {https://doi.org/10.1103/physrevb.83.155208} {\bibfield  {journal} {\bibinfo
  {journal} {Physical Review B}\ }\textbf {\bibinfo {volume} {83}},\ \bibinfo
  {pages} {155208} (\bibinfo {year} {2011})}\BibitemShut {NoStop}%
\bibitem [{\citenamefont {Settels}\ \emph {et~al.}(2014)\citenamefont
  {Settels}, \citenamefont {Schubert}, \citenamefont {Tafipolski},
  \citenamefont {Liu}, \citenamefont {Stehr}, \citenamefont {Topczak},
  \citenamefont {Pflaum}, \citenamefont {Deibel}, \citenamefont {Fink},
  \citenamefont {Engel},\ and\ \citenamefont {Engels}}]{iourpaamrfiefipbos}%
  \BibitemOpen
  \bibfield  {author} {\bibinfo {author} {\bibfnamefont {V.}~\bibnamefont
  {Settels}}, \bibinfo {author} {\bibfnamefont {A.}~\bibnamefont {Schubert}},
  \bibinfo {author} {\bibfnamefont {M.}~\bibnamefont {Tafipolski}}, \bibinfo
  {author} {\bibfnamefont {W.}~\bibnamefont {Liu}}, \bibinfo {author}
  {\bibfnamefont {V.}~\bibnamefont {Stehr}}, \bibinfo {author} {\bibfnamefont
  {A.~K.}\ \bibnamefont {Topczak}}, \bibinfo {author} {\bibfnamefont
  {J.}~\bibnamefont {Pflaum}}, \bibinfo {author} {\bibfnamefont
  {C.}~\bibnamefont {Deibel}}, \bibinfo {author} {\bibfnamefont {R.~F.}\
  \bibnamefont {Fink}}, \bibinfo {author} {\bibfnamefont {V.}~\bibnamefont
  {Engel}},\ and\ \bibinfo {author} {\bibfnamefont {B.}~\bibnamefont
  {Engels}},\ }\bibfield  {title} {\enquote {\bibinfo {title} {Identification
  of ultrafast relaxation processes as a major reason for inefficient exciton
  diffusion in perylene-based organic semiconductors},}\ }\href
  {https://doi.org/10.1021/ja413115h} {\bibfield  {journal} {\bibinfo
  {journal} {Journal of the American Chemical Society}\ }\textbf {\bibinfo
  {volume} {136}},\ \bibinfo {pages} {9327--9337} (\bibinfo {year}
  {2014})}\BibitemShut {NoStop}%
\bibitem [{\citenamefont {Stehr}\ \emph {et~al.}(2014)\citenamefont {Stehr},
  \citenamefont {Fink}, \citenamefont {Engels}, \citenamefont {Pflaum},\ and\
  \citenamefont {Deibel}}]{sefiocbomtr}%
  \BibitemOpen
  \bibfield  {author} {\bibinfo {author} {\bibfnamefont {V.}~\bibnamefont
  {Stehr}}, \bibinfo {author} {\bibfnamefont {R.~F.}\ \bibnamefont {Fink}},
  \bibinfo {author} {\bibfnamefont {B.}~\bibnamefont {Engels}}, \bibinfo
  {author} {\bibfnamefont {J.}~\bibnamefont {Pflaum}},\ and\ \bibinfo {author}
  {\bibfnamefont {C.}~\bibnamefont {Deibel}},\ }\bibfield  {title} {\enquote
  {\bibinfo {title} {Singlet exciton diffusion in organic crystals based on
  marcus transfer rates},}\ }\href {https://doi.org/10.1021/ct500014h}
  {\bibfield  {journal} {\bibinfo  {journal} {Journal of Chemical Theory and
  Computation}\ }\textbf {\bibinfo {volume} {10}},\ \bibinfo {pages}
  {1242--1255} (\bibinfo {year} {2014})}\BibitemShut {NoStop}%
\bibitem [{\citenamefont {McDowall}, \citenamefont {Johnson},\ and\
  \citenamefont {Patrick}(2010)}]{solsceopafa}%
  \BibitemOpen
  \bibfield  {author} {\bibinfo {author} {\bibfnamefont {S.}~\bibnamefont
  {McDowall}}, \bibinfo {author} {\bibfnamefont {B.~L.}\ \bibnamefont
  {Johnson}},\ and\ \bibinfo {author} {\bibfnamefont {D.~L.}\ \bibnamefont
  {Patrick}},\ }\bibfield  {title} {\enquote {\bibinfo {title} {Simulations of
  luminescent solar concentrators: Effects of polarization and fluorophore
  alignment},}\ }\href {https://doi.org/10.1063/1.3467801} {\bibfield
  {journal} {\bibinfo  {journal} {Journal of Applied Physics}\ }\textbf
  {\bibinfo {volume} {108}},\ \bibinfo {pages} {053508} (\bibinfo {year}
  {2010})}\BibitemShut {NoStop}%
\bibitem [{\citenamefont {{\c{S}}ahin}, \citenamefont {Ilan},\ and\
  \citenamefont {Kelley}(2011)}]{mcsolpilscbosn}%
  \BibitemOpen
  \bibfield  {author} {\bibinfo {author} {\bibfnamefont {D.}~\bibnamefont
  {{\c{S}}ahin}}, \bibinfo {author} {\bibfnamefont {B.}~\bibnamefont {Ilan}},\
  and\ \bibinfo {author} {\bibfnamefont {D.~F.}\ \bibnamefont {Kelley}},\
  }\bibfield  {title} {\enquote {\bibinfo {title} {Monte-carlo simulations of
  light propagation in luminescent solar concentrators based on semiconductor
  nanoparticles},}\ }\href {https://doi.org/10.1063/1.3619809} {\bibfield
  {journal} {\bibinfo  {journal} {Journal of Applied Physics}\ }\textbf
  {\bibinfo {volume} {110}},\ \bibinfo {pages} {033108} (\bibinfo {year}
  {2011})}\BibitemShut {NoStop}%
\bibitem [{\citenamefont {Schüler}\ \emph {et~al.}(2008)\citenamefont
  {Schüler}, \citenamefont {Kostro}, \citenamefont {Galande}, \citenamefont
  {del Olmo}, \citenamefont {de~Chambrier},\ and\ \citenamefont
  {Huriet}}]{pomkrtsiqdsc}%
  \BibitemOpen
  \bibfield  {author} {\bibinfo {author} {\bibfnamefont {A.}~\bibnamefont
  {Schüler}}, \bibinfo {author} {\bibfnamefont {A.}~\bibnamefont {Kostro}},
  \bibinfo {author} {\bibfnamefont {C.}~\bibnamefont {Galande}}, \bibinfo
  {author} {\bibfnamefont {M.~V.}\ \bibnamefont {del Olmo}}, \bibinfo {author}
  {\bibfnamefont {E.}~\bibnamefont {de~Chambrier}},\ and\ \bibinfo {author}
  {\bibfnamefont {B.}~\bibnamefont {Huriet}},\ }\bibfield  {title} {\enquote
  {\bibinfo {title} {Principles of monte-carlo ray-tracing simulations of
  quantum dot solar concentrators},}\ }in\ \href
  {https://doi.org/10.1007/978-3-540-75997-3_200} {\emph {\bibinfo {booktitle}
  {Proceedings of {ISES} World Congress 2007 (Vol. I {\textendash} Vol. V)}}}\
  (\bibinfo  {publisher} {Springer Berlin Heidelberg},\ \bibinfo {year}
  {2008})\ pp.\ \bibinfo {pages} {1033--1037}\BibitemShut {NoStop}%
\bibitem [{\citenamefont {Andersen}, \citenamefont {Panosetti},\ and\
  \citenamefont {Reuter}(2019{\natexlab{a}})}]{apgtskmcs}%
  \BibitemOpen
  \bibfield  {author} {\bibinfo {author} {\bibfnamefont {M.}~\bibnamefont
  {Andersen}}, \bibinfo {author} {\bibfnamefont {C.}~\bibnamefont
  {Panosetti}},\ and\ \bibinfo {author} {\bibfnamefont {K.}~\bibnamefont
  {Reuter}},\ }\bibfield  {title} {\enquote {\bibinfo {title} {A practical
  guide to surface kinetic monte carlo simulations},}\ }\href
  {https://doi.org/10.3389/fchem.2019.00202} {\bibfield  {journal} {\bibinfo
  {journal} {Front. Chem.}\ }\textbf {\bibinfo {volume} {7}} (\bibinfo {year}
  {2019}{\natexlab{a}}),\ 10.3389/fchem.2019.00202}\BibitemShut {NoStop}%
\bibitem [{\citenamefont {de~Sousa}\ \emph {et~al.}(2018)\citenamefont
  {de~Sousa}, \citenamefont {da~Silva~Filho}, \citenamefont {de~Sousa},\ and\
  \citenamefont {de~Oliveira~Neto}}]{edionfamcsoteotad}%
  \BibitemOpen
  \bibfield  {author} {\bibinfo {author} {\bibfnamefont {L.~E.}\ \bibnamefont
  {de~Sousa}}, \bibinfo {author} {\bibfnamefont {D.~A.}\ \bibnamefont
  {da~Silva~Filho}}, \bibinfo {author} {\bibfnamefont {R.~T.}\ \bibnamefont
  {de~Sousa}},\ and\ \bibinfo {author} {\bibfnamefont {P.~H.}\ \bibnamefont
  {de~Oliveira~Neto}},\ }\bibfield  {title} {\enquote {\bibinfo {title}
  {Exciton diffusion in organic nanofibers: A monte carlo study on the effects
  of temperature and dimensionality},}\ }\href
  {https://doi.org/10.1038/s41598-018-32232-5} {\bibfield  {journal} {\bibinfo
  {journal} {Sci. Rep.}\ }\textbf {\bibinfo {volume} {8}} (\bibinfo {year}
  {2018}),\ 10.1038/s41598-018-32232-5}\BibitemShut {NoStop}%
\bibitem [{\citenamefont {Suarez}, \citenamefont {Menger},\ and\ \citenamefont
  {Faraji}(2020)}]{sfitaesa}%
  \BibitemOpen
  \bibfield  {author} {\bibinfo {author} {\bibfnamefont {L.~E.~A.}\
  \bibnamefont {Suarez}}, \bibinfo {author} {\bibfnamefont {M.~F. S.~J.}\
  \bibnamefont {Menger}},\ and\ \bibinfo {author} {\bibfnamefont
  {S.}~\bibnamefont {Faraji}},\ }\bibfield  {title} {\enquote {\bibinfo {title}
  {Singlet fission in tetracene: an excited state analysis},}\ }\href
  {https://doi.org/10.1080/00268976.2020.1769870} {\bibfield  {journal}
  {\bibinfo  {journal} {Mol Phys}\ }\textbf {\bibinfo {volume} {118}},\
  \bibinfo {pages} {e1769870} (\bibinfo {year} {2020})}\BibitemShut {NoStop}%
\bibitem [{\citenamefont {Burdett}\ and\ \citenamefont
  {Bardeen}(2013)}]{tdosfictaca}%
  \BibitemOpen
  \bibfield  {author} {\bibinfo {author} {\bibfnamefont {J.~J.}\ \bibnamefont
  {Burdett}}\ and\ \bibinfo {author} {\bibfnamefont {C.~J.}\ \bibnamefont
  {Bardeen}},\ }\bibfield  {title} {\enquote {\bibinfo {title} {The dynamics of
  singlet fission in crystalline tetracene and covalent analogs},}\ }\href
  {https://doi.org/10.1021/ar300191w} {\bibfield  {journal} {\bibinfo
  {journal} {Acc. Chem. Res.}\ }\textbf {\bibinfo {volume} {46}},\ \bibinfo
  {pages} {1312--1320} (\bibinfo {year} {2013})}\BibitemShut {NoStop}%
\bibitem [{\citenamefont {Burdett}\ \emph {et~al.}(2010)\citenamefont
  {Burdett}, \citenamefont {Müller}, \citenamefont {Gosztola},\ and\
  \citenamefont {Bardeen}}]{esdisamttrosraef}%
  \BibitemOpen
  \bibfield  {author} {\bibinfo {author} {\bibfnamefont {J.~J.}\ \bibnamefont
  {Burdett}}, \bibinfo {author} {\bibfnamefont {A.~M.}\ \bibnamefont
  {Müller}}, \bibinfo {author} {\bibfnamefont {D.}~\bibnamefont {Gosztola}},\
  and\ \bibinfo {author} {\bibfnamefont {C.~J.}\ \bibnamefont {Bardeen}},\
  }\bibfield  {title} {\enquote {\bibinfo {title} {Excited state dynamics in
  solid and monomeric tetracene: The roles of superradiance and exciton
  fission},}\ }\href {https://doi.org/10.1063/1.3495764} {\bibfield  {journal}
  {\bibinfo  {journal} {The Journal of Chemical Physics}\ }\textbf {\bibinfo
  {volume} {133}},\ \bibinfo {pages} {144506} (\bibinfo {year}
  {2010})}\BibitemShut {NoStop}%
\bibitem [{\citenamefont {Br{\'{e}}das}\ \emph {et~al.}(2009)\citenamefont
  {Br{\'{e}}das}, \citenamefont {Norton}, \citenamefont {Cornil},\ and\
  \citenamefont {Coropceanu}}]{muoosc}%
  \BibitemOpen
  \bibfield  {author} {\bibinfo {author} {\bibfnamefont {J.-L.}\ \bibnamefont
  {Br{\'{e}}das}}, \bibinfo {author} {\bibfnamefont {J.~E.}\ \bibnamefont
  {Norton}}, \bibinfo {author} {\bibfnamefont {J.}~\bibnamefont {Cornil}},\
  and\ \bibinfo {author} {\bibfnamefont {V.}~\bibnamefont {Coropceanu}},\
  }\bibfield  {title} {\enquote {\bibinfo {title} {Molecular understanding of
  organic solar cells: The challenges},}\ }\href
  {https://doi.org/10.1021/ar900099h} {\bibfield  {journal} {\bibinfo
  {journal} {Accounts of Chemical Research}\ }\textbf {\bibinfo {volume}
  {42}},\ \bibinfo {pages} {1691--1699} (\bibinfo {year} {2009})}\BibitemShut
  {NoStop}%
\bibitem [{\citenamefont {Balzer}\ \emph {et~al.}(2021)\citenamefont {Balzer},
  \citenamefont {Smolders}, \citenamefont {Blyth}, \citenamefont {Hood},\ and\
  \citenamefont {Kassal}}]{dkmcfsdecaetidm}%
  \BibitemOpen
  \bibfield  {author} {\bibinfo {author} {\bibfnamefont {D.}~\bibnamefont
  {Balzer}}, \bibinfo {author} {\bibfnamefont {T.~J. A.~M.}\ \bibnamefont
  {Smolders}}, \bibinfo {author} {\bibfnamefont {D.}~\bibnamefont {Blyth}},
  \bibinfo {author} {\bibfnamefont {S.~N.}\ \bibnamefont {Hood}},\ and\
  \bibinfo {author} {\bibfnamefont {I.}~\bibnamefont {Kassal}},\ }\bibfield
  {title} {\enquote {\bibinfo {title} {Delocalised kinetic monte carlo for
  simulating delocalisation-enhanced charge and exciton transport in disordered
  materials},}\ }\href {https://doi.org/10.1039/d0sc04116e} {\bibfield
  {journal} {\bibinfo  {journal} {Chemical Science}\ }\textbf {\bibinfo
  {volume} {12}},\ \bibinfo {pages} {2276--2285} (\bibinfo {year}
  {2021})}\BibitemShut {NoStop}%
\bibitem [{\citenamefont {F\"{o}rster}(1948)}]{forster}%
  \BibitemOpen
  \bibfield  {author} {\bibinfo {author} {\bibfnamefont {T.}~\bibnamefont
  {F\"{o}rster}},\ }\bibfield  {title} {\enquote {\bibinfo {title} {Energy
  migration and fluorescence},}\ }\href
  {https://doi.org/10.1002/andp.19484370105} {\bibfield  {journal} {\bibinfo
  {journal} {J Biomed Opt}\ }\textbf {\bibinfo {volume} {437}},\ \bibinfo
  {pages} {55--75} (\bibinfo {year} {1948})}\BibitemShut {NoStop}%
\bibitem [{\citenamefont {Dexter}(1953)}]{dexter}%
  \BibitemOpen
  \bibfield  {author} {\bibinfo {author} {\bibfnamefont {D.~L.}\ \bibnamefont
  {Dexter}},\ }\bibfield  {title} {\enquote {\bibinfo {title} {A theory of
  sensitized luminescence in solids},}\ }\href
  {https://doi.org/10.1063/1.1699044} {\bibfield  {journal} {\bibinfo
  {journal} {J Chem Phys}\ }\textbf {\bibinfo {volume} {21}},\ \bibinfo {pages}
  {836--850} (\bibinfo {year} {1953})}\BibitemShut {NoStop}%
\bibitem [{\citenamefont {Banerjee}\ \emph {et~al.}(2016)\citenamefont
  {Banerjee}, \citenamefont {Baiardi}, \citenamefont {Bloino},\ and\
  \citenamefont {Barone}}]{veoroeetattd}%
  \BibitemOpen
  \bibfield  {author} {\bibinfo {author} {\bibfnamefont {S.}~\bibnamefont
  {Banerjee}}, \bibinfo {author} {\bibfnamefont {A.}~\bibnamefont {Baiardi}},
  \bibinfo {author} {\bibfnamefont {J.}~\bibnamefont {Bloino}},\ and\ \bibinfo
  {author} {\bibfnamefont {V.}~\bibnamefont {Barone}},\ }\bibfield  {title}
  {\enquote {\bibinfo {title} {Vibronic effects on rates of excitation energy
  transfer and their temperature dependence},}\ }\href
  {https://doi.org/10.1021/acs.jctc.6b00157} {\bibfield  {journal} {\bibinfo
  {journal} {J Chem Theory Comput}\ }\textbf {\bibinfo {volume} {12}},\
  \bibinfo {pages} {2357--2365} (\bibinfo {year} {2016})}\BibitemShut {NoStop}%
\bibitem [{\citenamefont {Kasha}(1950)}]{kasha}%
  \BibitemOpen
  \bibfield  {author} {\bibinfo {author} {\bibfnamefont {M.}~\bibnamefont
  {Kasha}},\ }\bibfield  {title} {\enquote {\bibinfo {title} {Characterization
  of electronic transitions in complex molecules},}\ }\href@noop {} {\bibfield
  {journal} {\bibinfo  {journal} {Discuss Faraday Soc}\ ,\ \bibinfo {pages}
  {14--19}} (\bibinfo {year} {1950})}\BibitemShut {NoStop}%
\bibitem [{\citenamefont {Voter}(2007)}]{Voter2007}%
  \BibitemOpen
  \bibfield  {author} {\bibinfo {author} {\bibfnamefont {A.~F.}\ \bibnamefont
  {Voter}},\ }\bibfield  {title} {\enquote {\bibinfo {title} {{Introduction To
  The Kinetic Monte Carlo Method}},}\ }\href
  {https://doi.org/10.1007/978-1-4020-5295-8_1} {\bibfield  {journal} {\bibinfo
   {journal} {Radiation Effects in Solids}\ ,\ \bibinfo {pages} {1--23}}
  (\bibinfo {year} {2007})}\BibitemShut {NoStop}%
\bibitem [{\citenamefont {Stamatakis}\ and\ \citenamefont
  {Vlachos}(2011)}]{Stamatakis2011}%
  \BibitemOpen
  \bibfield  {author} {\bibinfo {author} {\bibfnamefont {M.}~\bibnamefont
  {Stamatakis}}\ and\ \bibinfo {author} {\bibfnamefont {D.~G.}\ \bibnamefont
  {Vlachos}},\ }\bibfield  {title} {\enquote {\bibinfo {title} {{A
  graph-theoretical kinetic Monte Carlo framework for on-lattice chemical
  kinetics}},}\ }\href {https://doi.org/10.1063/1.3596751} {\bibfield
  {journal} {\bibinfo  {journal} {The Journal of Chemical Physics}\ }\textbf
  {\bibinfo {volume} {134}},\ \bibinfo {pages} {214115} (\bibinfo {year}
  {2011})}\BibitemShut {NoStop}%
\bibitem [{\citenamefont {Walker}, \citenamefont {Vogt},\ and\ \citenamefont
  {Cole}(2015)}]{Walker2015}%
  \BibitemOpen
  \bibfield  {author} {\bibinfo {author} {\bibfnamefont {K.~A.}\ \bibnamefont
  {Walker}}, \bibinfo {author} {\bibfnamefont {N.}~\bibnamefont {Vogt}},\ and\
  \bibinfo {author} {\bibfnamefont {J.~H.}\ \bibnamefont {Cole}},\ }\bibfield
  {title} {\enquote {\bibinfo {title} {{Charge filling factors in clean and
  disordered arrays of tunnel junctions}},}\ }\href
  {https://doi.org/10.1038/srep17572} {\bibfield  {journal} {\bibinfo
  {journal} {Scientific Reports 2015 5:1}\ }\textbf {\bibinfo {volume} {5}},\
  \bibinfo {pages} {1--8} (\bibinfo {year} {2015})}\BibitemShut {NoStop}%
\bibitem [{\citenamefont {Ostroverkhova}(2016)}]{Ostroverkhova2016}%
  \BibitemOpen
  \bibfield  {author} {\bibinfo {author} {\bibfnamefont {O.}~\bibnamefont
  {Ostroverkhova}},\ }\bibfield  {title} {\enquote {\bibinfo {title} {{Organic
  Optoelectronic Materials: Mechanisms and Applications}},}\ }\href
  {https://doi.org/10.1021/acs.chemrev.6b00127} {\bibfield  {journal} {\bibinfo
   {journal} {Chemical Reviews}\ }\textbf {\bibinfo {volume} {116}},\ \bibinfo
  {pages} {13279--13412} (\bibinfo {year} {2016})}\BibitemShut {NoStop}%
\bibitem [{\citenamefont {Oberhofer}, \citenamefont {Reuter},\ and\
  \citenamefont {Blumberger}(2017)}]{Oberhofer2017}%
  \BibitemOpen
  \bibfield  {author} {\bibinfo {author} {\bibfnamefont {H.}~\bibnamefont
  {Oberhofer}}, \bibinfo {author} {\bibfnamefont {K.}~\bibnamefont {Reuter}},\
  and\ \bibinfo {author} {\bibfnamefont {J.}~\bibnamefont {Blumberger}},\
  }\bibfield  {title} {\enquote {\bibinfo {title} {{Charge Transport in
  Molecular Materials: An Assessment of Computational Methods}},}\ }\href
  {https://doi.org/10.1021/acs.chemrev.7b00086} {\bibfield  {journal} {\bibinfo
   {journal} {Chemical Reviews}\ }\textbf {\bibinfo {volume} {117}},\ \bibinfo
  {pages} {10319--10357} (\bibinfo {year} {2017})}\BibitemShut {NoStop}%
\bibitem [{\citenamefont {Lin}, \citenamefont {Huang},\ and\ \citenamefont
  {Hin}(2018)}]{Lin2018}%
  \BibitemOpen
  \bibfield  {author} {\bibinfo {author} {\bibfnamefont {F.}~\bibnamefont
  {Lin}}, \bibinfo {author} {\bibfnamefont {J.}~\bibnamefont {Huang}},\ and\
  \bibinfo {author} {\bibfnamefont {C.}~\bibnamefont {Hin}},\ }\bibfield
  {title} {\enquote {\bibinfo {title} {{Electron Transport from Quantum Kinetic
  Monte Carlo Simulations}},}\ }\href
  {https://doi.org/10.1021/acs.jpcc.8b05347} {\bibfield  {journal} {\bibinfo
  {journal} {Journal of Physical Chemistry C}\ }\textbf {\bibinfo {volume}
  {122}},\ \bibinfo {pages} {20550--20554} (\bibinfo {year} {2018})},\ \Eprint
  {https://arxiv.org/abs/1704.07545} {arXiv:1704.07545} \BibitemShut {NoStop}%
\bibitem [{\citenamefont {J{\o}rgensen}\ and\ \citenamefont
  {Gr{\"{o}}nbeck}(2018)}]{Jorgensen2018}%
  \BibitemOpen
  \bibfield  {author} {\bibinfo {author} {\bibfnamefont {M.}~\bibnamefont
  {J{\o}rgensen}}\ and\ \bibinfo {author} {\bibfnamefont {H.}~\bibnamefont
  {Gr{\"{o}}nbeck}},\ }\bibfield  {title} {\enquote {\bibinfo {title}
  {{MonteCoffee: A programmable kinetic Monte Carlo framework}},}\ }\href
  {https://doi.org/10.1063/1.5046635} {\bibfield  {journal} {\bibinfo
  {journal} {The Journal of Chemical Physics}\ }\textbf {\bibinfo {volume}
  {149}},\ \bibinfo {pages} {114101} (\bibinfo {year} {2018})}\BibitemShut
  {NoStop}%
\bibitem [{\citenamefont {Andersen}, \citenamefont {Panosetti},\ and\
  \citenamefont {Reuter}(2019{\natexlab{b}})}]{Andersen2019}%
  \BibitemOpen
  \bibfield  {author} {\bibinfo {author} {\bibfnamefont {M.}~\bibnamefont
  {Andersen}}, \bibinfo {author} {\bibfnamefont {C.}~\bibnamefont
  {Panosetti}},\ and\ \bibinfo {author} {\bibfnamefont {K.}~\bibnamefont
  {Reuter}},\ }\href {https://doi.org/10.3389/fchem.2019.00202} {\enquote
  {\bibinfo {title} {{A practical guide to surface kinetic Monte Carlo
  simulations}},}\ } (\bibinfo {year} {2019}{\natexlab{b}}),\ \Eprint
  {https://arxiv.org/abs/1904.02561} {arXiv:1904.02561} \BibitemShut {NoStop}%
\bibitem [{\citenamefont {Rego}\ and\ \citenamefont
  {Brandao}(2020)}]{Rego2020}%
  \BibitemOpen
  \bibfield  {author} {\bibinfo {author} {\bibfnamefont {A.~S.}\ \bibnamefont
  {Rego}}\ and\ \bibinfo {author} {\bibfnamefont {A.~L.}\ \bibnamefont
  {Brandao}},\ }\bibfield  {title} {\enquote {\bibinfo {title} {{General method
  for speeding up kinetic monte carlo simulations}},}\ }\href
  {https://doi.org/10.1021/acs.iecr.0c01069} {\bibfield  {journal} {\bibinfo
  {journal} {Industrial and Engineering Chemistry Research}\ }\textbf {\bibinfo
  {volume} {59}},\ \bibinfo {pages} {9034--9042} (\bibinfo {year}
  {2020})}\BibitemShut {NoStop}%
\bibitem [{\citenamefont {Vogt}, \citenamefont {Jeske},\ and\ \citenamefont
  {Cole}(2013)}]{sbrtqjiasse}%
  \BibitemOpen
  \bibfield  {author} {\bibinfo {author} {\bibfnamefont {N.}~\bibnamefont
  {Vogt}}, \bibinfo {author} {\bibfnamefont {J.}~\bibnamefont {Jeske}},\ and\
  \bibinfo {author} {\bibfnamefont {J.~H.}\ \bibnamefont {Cole}},\ }\bibfield
  {title} {\enquote {\bibinfo {title} {Stochastic bloch-redfield theory:
  Quantum jumps in a solid-state environment},}\ }\href
  {https://doi.org/10.1103/physrevb.88.174514} {\bibfield  {journal} {\bibinfo
  {journal} {Physical Review B}\ }\textbf {\bibinfo {volume} {88}},\ \bibinfo
  {pages} {174514} (\bibinfo {year} {2013})}\BibitemShut {NoStop}%
\bibitem [{\citenamefont {Kranz}\ and\ \citenamefont
  {Elstner}(2016)}]{Kranz2016a}%
  \BibitemOpen
  \bibfield  {author} {\bibinfo {author} {\bibfnamefont {J.~J.}\ \bibnamefont
  {Kranz}}\ and\ \bibinfo {author} {\bibfnamefont {M.}~\bibnamefont
  {Elstner}},\ }\bibfield  {title} {\enquote {\bibinfo {title} {{Simulation of
  Singlet Exciton Diffusion in Bulk Organic Materials}},}\ }\href
  {https://doi.org/10.1021/acs.jctc.6b00235} {\bibfield  {journal} {\bibinfo
  {journal} {Journal of Chemical Theory and Computation}\ }\textbf {\bibinfo
  {volume} {12}},\ \bibinfo {pages} {4209--4221} (\bibinfo {year}
  {2016})}\BibitemShut {NoStop}%
\bibitem [{\citenamefont {Lyskov}, \citenamefont {Kleinschmidt},\ and\
  \citenamefont {Marian}(2016)}]{MRCI_igor}%
  \BibitemOpen
  \bibfield  {author} {\bibinfo {author} {\bibfnamefont {I.}~\bibnamefont
  {Lyskov}}, \bibinfo {author} {\bibfnamefont {M.}~\bibnamefont
  {Kleinschmidt}},\ and\ \bibinfo {author} {\bibfnamefont {C.~M.}\ \bibnamefont
  {Marian}},\ }\bibfield  {title} {\enquote {\bibinfo {title} {Redesign of the
  {DFT}/{MRCI} hamiltonian},}\ }\href {https://doi.org/10.1063/1.4940036}
  {\bibfield  {journal} {\bibinfo  {journal} {J Chem Phys}\ }\textbf {\bibinfo
  {volume} {144}},\ \bibinfo {pages} {034104} (\bibinfo {year}
  {2016})}\BibitemShut {NoStop}%
\bibitem [{\citenamefont {Grell}\ and\ \citenamefont {Bradley}(1999)}]{plfomm}%
  \BibitemOpen
  \bibfield  {author} {\bibinfo {author} {\bibfnamefont {M.}~\bibnamefont
  {Grell}}\ and\ \bibinfo {author} {\bibfnamefont {D.~D.~C.}\ \bibnamefont
  {Bradley}},\ }\bibfield  {title} {\enquote {\bibinfo {title} {Polarized
  luminescence from oriented molecular materials},}\ }\href
  {https://doi.org/10.1002/(sici)1521-4095(199908)11:11<895::aid-adma895>3.0.co;2-y}
  {\bibfield  {journal} {\bibinfo  {journal} {Advanced Materials}\ }\textbf
  {\bibinfo {volume} {11}},\ \bibinfo {pages} {895--905} (\bibinfo {year}
  {1999})}\BibitemShut {NoStop}%
\bibitem [{\citenamefont {Lunt}\ \emph {et~al.}(2009)\citenamefont {Lunt},
  \citenamefont {Giebink}, \citenamefont {Belak}, \citenamefont {Benziger},\
  and\ \citenamefont {Forrest}}]{edloostfmbsrplq}%
  \BibitemOpen
  \bibfield  {author} {\bibinfo {author} {\bibfnamefont {R.~R.}\ \bibnamefont
  {Lunt}}, \bibinfo {author} {\bibfnamefont {N.~C.}\ \bibnamefont {Giebink}},
  \bibinfo {author} {\bibfnamefont {A.~A.}\ \bibnamefont {Belak}}, \bibinfo
  {author} {\bibfnamefont {J.~B.}\ \bibnamefont {Benziger}},\ and\ \bibinfo
  {author} {\bibfnamefont {S.~R.}\ \bibnamefont {Forrest}},\ }\bibfield
  {title} {\enquote {\bibinfo {title} {Exciton diffusion lengths of organic
  semiconductor thin films measured by spectrally resolved photoluminescence
  quenching},}\ }\href {https://doi.org/10.1063/1.3079797} {\bibfield
  {journal} {\bibinfo  {journal} {Journal of Applied Physics}\ }\textbf
  {\bibinfo {volume} {105}},\ \bibinfo {pages} {053711} (\bibinfo {year}
  {2009})}\BibitemShut {NoStop}%
\bibitem [{\citenamefont {Becke}(1993)}]{b3lyp1}%
  \BibitemOpen
  \bibfield  {author} {\bibinfo {author} {\bibfnamefont {A.~D.}\ \bibnamefont
  {Becke}},\ }\bibfield  {title} {\enquote {\bibinfo {title}
  {Density-functional thermochemistry. {III}. the role of exact exchange},}\
  }\href {https://doi.org/10.1063/1.464913} {\bibfield  {journal} {\bibinfo
  {journal} {J Chem Phys}\ }\textbf {\bibinfo {volume} {98}},\ \bibinfo {pages}
  {5648--5652} (\bibinfo {year} {1993})}\BibitemShut {NoStop}%
\bibitem [{\citenamefont {Schäfer}, \citenamefont {Huber},\ and\ \citenamefont
  {Ahlrichs}(1994)}]{TZVP}%
  \BibitemOpen
  \bibfield  {author} {\bibinfo {author} {\bibfnamefont {A.}~\bibnamefont
  {Schäfer}}, \bibinfo {author} {\bibfnamefont {C.}~\bibnamefont {Huber}},\
  and\ \bibinfo {author} {\bibfnamefont {R.}~\bibnamefont {Ahlrichs}},\
  }\bibfield  {title} {\enquote {\bibinfo {title} {Fully optimized contracted
  gaussian basis sets of triple zeta valence quality for atoms li to kr},}\
  }\href {https://doi.org/10.1063/1.467146} {\bibfield  {journal} {\bibinfo
  {journal} {J Chem Phys}\ }\textbf {\bibinfo {volume} {100}},\ \bibinfo
  {pages} {5829--5835} (\bibinfo {year} {1994})}\BibitemShut {NoStop}%
\bibitem [{\citenamefont {Frisch}\ \emph {et~al.}(2016)\citenamefont {Frisch},
  \citenamefont {Trucks}, \citenamefont {Schlegel}, \citenamefont {Scuseria},
  \citenamefont {Robb}, \citenamefont {Cheeseman}, \citenamefont {Scalmani},
  \citenamefont {Barone}, \citenamefont {Petersson}, \citenamefont {Nakatsuji},
  \citenamefont {Li}, \citenamefont {Caricato}, \citenamefont {Marenich},
  \citenamefont {Bloino}, \citenamefont {Janesko}, \citenamefont {Gomperts},
  \citenamefont {Mennucci}, \citenamefont {Hratchian}, \citenamefont {Ortiz},
  \citenamefont {Izmaylov}, \citenamefont {Sonnenberg}, \citenamefont
  {Williams-Young}, \citenamefont {Ding}, \citenamefont {Lipparini},
  \citenamefont {Egidi}, \citenamefont {Goings}, \citenamefont {Peng},
  \citenamefont {Petrone}, \citenamefont {Henderson}, \citenamefont
  {Ranasinghe}, \citenamefont {Zakrzewski}, \citenamefont {Gao}, \citenamefont
  {Rega}, \citenamefont {Zheng}, \citenamefont {Liang}, \citenamefont {Hada},
  \citenamefont {Ehara}, \citenamefont {Toyota}, \citenamefont {Fukuda},
  \citenamefont {Hasegawa}, \citenamefont {Ishida}, \citenamefont {Nakajima},
  \citenamefont {Honda}, \citenamefont {Kitao}, \citenamefont {Nakai},
  \citenamefont {Vreven}, \citenamefont {Throssell}, \citenamefont
  {Montgomery}, \citenamefont {Peralta}, \citenamefont {Ogliaro}, \citenamefont
  {Bearpark}, \citenamefont {Heyd}, \citenamefont {Brothers}, \citenamefont
  {Kudin}, \citenamefont {Staroverov}, \citenamefont {Keith}, \citenamefont
  {Kobayashi}, \citenamefont {Normand}, \citenamefont {Raghavachari},
  \citenamefont {Rendell}, \citenamefont {Burant}, \citenamefont {Iyengar},
  \citenamefont {Tomasi}, \citenamefont {Cossi}, \citenamefont {Millam},
  \citenamefont {Klene}, \citenamefont {Adamo}, \citenamefont {Cammi},
  \citenamefont {Ochterski}, \citenamefont {Martin}, \citenamefont {Morokuma},
  \citenamefont {Farkas}, \citenamefont {Foresman},\ and\ \citenamefont
  {Fox}}]{GAUSSIAN}%
  \BibitemOpen
  \bibfield  {author} {\bibinfo {author} {\bibfnamefont {M.~J.}\ \bibnamefont
  {Frisch}}, \bibinfo {author} {\bibfnamefont {G.~W.}\ \bibnamefont {Trucks}},
  \bibinfo {author} {\bibfnamefont {H.~B.}\ \bibnamefont {Schlegel}}, \bibinfo
  {author} {\bibfnamefont {G.~E.}\ \bibnamefont {Scuseria}}, \bibinfo {author}
  {\bibfnamefont {M.~A.}\ \bibnamefont {Robb}}, \bibinfo {author}
  {\bibfnamefont {J.~R.}\ \bibnamefont {Cheeseman}}, \bibinfo {author}
  {\bibfnamefont {G.}~\bibnamefont {Scalmani}}, \bibinfo {author}
  {\bibfnamefont {V.}~\bibnamefont {Barone}}, \bibinfo {author} {\bibfnamefont
  {G.~A.}\ \bibnamefont {Petersson}}, \bibinfo {author} {\bibfnamefont
  {H.}~\bibnamefont {Nakatsuji}}, \bibinfo {author} {\bibfnamefont
  {X.}~\bibnamefont {Li}}, \bibinfo {author} {\bibfnamefont {M.}~\bibnamefont
  {Caricato}}, \bibinfo {author} {\bibfnamefont {A.~V.}\ \bibnamefont
  {Marenich}}, \bibinfo {author} {\bibfnamefont {J.}~\bibnamefont {Bloino}},
  \bibinfo {author} {\bibfnamefont {B.~G.}\ \bibnamefont {Janesko}}, \bibinfo
  {author} {\bibfnamefont {R.}~\bibnamefont {Gomperts}}, \bibinfo {author}
  {\bibfnamefont {B.}~\bibnamefont {Mennucci}}, \bibinfo {author}
  {\bibfnamefont {H.~P.}\ \bibnamefont {Hratchian}}, \bibinfo {author}
  {\bibfnamefont {J.~V.}\ \bibnamefont {Ortiz}}, \bibinfo {author}
  {\bibfnamefont {A.~F.}\ \bibnamefont {Izmaylov}}, \bibinfo {author}
  {\bibfnamefont {J.~L.}\ \bibnamefont {Sonnenberg}}, \bibinfo {author}
  {\bibfnamefont {D.}~\bibnamefont {Williams-Young}}, \bibinfo {author}
  {\bibfnamefont {F.}~\bibnamefont {Ding}}, \bibinfo {author} {\bibfnamefont
  {F.}~\bibnamefont {Lipparini}}, \bibinfo {author} {\bibfnamefont
  {F.}~\bibnamefont {Egidi}}, \bibinfo {author} {\bibfnamefont
  {J.}~\bibnamefont {Goings}}, \bibinfo {author} {\bibfnamefont
  {B.}~\bibnamefont {Peng}}, \bibinfo {author} {\bibfnamefont {A.}~\bibnamefont
  {Petrone}}, \bibinfo {author} {\bibfnamefont {T.}~\bibnamefont {Henderson}},
  \bibinfo {author} {\bibfnamefont {D.}~\bibnamefont {Ranasinghe}}, \bibinfo
  {author} {\bibfnamefont {V.~G.}\ \bibnamefont {Zakrzewski}}, \bibinfo
  {author} {\bibfnamefont {J.}~\bibnamefont {Gao}}, \bibinfo {author}
  {\bibfnamefont {N.}~\bibnamefont {Rega}}, \bibinfo {author} {\bibfnamefont
  {G.}~\bibnamefont {Zheng}}, \bibinfo {author} {\bibfnamefont
  {W.}~\bibnamefont {Liang}}, \bibinfo {author} {\bibfnamefont
  {M.}~\bibnamefont {Hada}}, \bibinfo {author} {\bibfnamefont {M.}~\bibnamefont
  {Ehara}}, \bibinfo {author} {\bibfnamefont {K.}~\bibnamefont {Toyota}},
  \bibinfo {author} {\bibfnamefont {R.}~\bibnamefont {Fukuda}}, \bibinfo
  {author} {\bibfnamefont {J.}~\bibnamefont {Hasegawa}}, \bibinfo {author}
  {\bibfnamefont {M.}~\bibnamefont {Ishida}}, \bibinfo {author} {\bibfnamefont
  {T.}~\bibnamefont {Nakajima}}, \bibinfo {author} {\bibfnamefont
  {Y.}~\bibnamefont {Honda}}, \bibinfo {author} {\bibfnamefont
  {O.}~\bibnamefont {Kitao}}, \bibinfo {author} {\bibfnamefont
  {H.}~\bibnamefont {Nakai}}, \bibinfo {author} {\bibfnamefont
  {T.}~\bibnamefont {Vreven}}, \bibinfo {author} {\bibfnamefont
  {K.}~\bibnamefont {Throssell}}, \bibinfo {author} {\bibfnamefont {J.~A.}\
  \bibnamefont {Montgomery}, \bibfnamefont {{Jr.}}}, \bibinfo {author}
  {\bibfnamefont {J.~E.}\ \bibnamefont {Peralta}}, \bibinfo {author}
  {\bibfnamefont {F.}~\bibnamefont {Ogliaro}}, \bibinfo {author} {\bibfnamefont
  {M.~J.}\ \bibnamefont {Bearpark}}, \bibinfo {author} {\bibfnamefont {J.~J.}\
  \bibnamefont {Heyd}}, \bibinfo {author} {\bibfnamefont {E.~N.}\ \bibnamefont
  {Brothers}}, \bibinfo {author} {\bibfnamefont {K.~N.}\ \bibnamefont {Kudin}},
  \bibinfo {author} {\bibfnamefont {V.~N.}\ \bibnamefont {Staroverov}},
  \bibinfo {author} {\bibfnamefont {T.~A.}\ \bibnamefont {Keith}}, \bibinfo
  {author} {\bibfnamefont {R.}~\bibnamefont {Kobayashi}}, \bibinfo {author}
  {\bibfnamefont {J.}~\bibnamefont {Normand}}, \bibinfo {author} {\bibfnamefont
  {K.}~\bibnamefont {Raghavachari}}, \bibinfo {author} {\bibfnamefont {A.~P.}\
  \bibnamefont {Rendell}}, \bibinfo {author} {\bibfnamefont {J.~C.}\
  \bibnamefont {Burant}}, \bibinfo {author} {\bibfnamefont {S.~S.}\
  \bibnamefont {Iyengar}}, \bibinfo {author} {\bibfnamefont {J.}~\bibnamefont
  {Tomasi}}, \bibinfo {author} {\bibfnamefont {M.}~\bibnamefont {Cossi}},
  \bibinfo {author} {\bibfnamefont {J.~M.}\ \bibnamefont {Millam}}, \bibinfo
  {author} {\bibfnamefont {M.}~\bibnamefont {Klene}}, \bibinfo {author}
  {\bibfnamefont {C.}~\bibnamefont {Adamo}}, \bibinfo {author} {\bibfnamefont
  {R.}~\bibnamefont {Cammi}}, \bibinfo {author} {\bibfnamefont {J.~W.}\
  \bibnamefont {Ochterski}}, \bibinfo {author} {\bibfnamefont {R.~L.}\
  \bibnamefont {Martin}}, \bibinfo {author} {\bibfnamefont {K.}~\bibnamefont
  {Morokuma}}, \bibinfo {author} {\bibfnamefont {O.}~\bibnamefont {Farkas}},
  \bibinfo {author} {\bibfnamefont {J.~B.}\ \bibnamefont {Foresman}},\ and\
  \bibinfo {author} {\bibfnamefont {D.~J.}\ \bibnamefont {Fox}},\ }\href@noop
  {} {\enquote {\bibinfo {title} {Gaussian16 {R}evision {B}.01},}\ } (\bibinfo
  {year} {2016}),\ \bibinfo {note} {gaussian Inc. Wallingford CT}\BibitemShut
  {NoStop}%
\bibitem [{\citenamefont {Becke}(1988)}]{bhlyp}%
  \BibitemOpen
  \bibfield  {author} {\bibinfo {author} {\bibfnamefont {A.~D.}\ \bibnamefont
  {Becke}},\ }\bibfield  {title} {\enquote {\bibinfo {title}
  {Density-functional exchange-energy approximation with correct asymptotic
  behavior},}\ }\href {https://doi.org/10.1103/physreva.38.3098} {\bibfield
  {journal} {\bibinfo  {journal} {Phys Rev A}\ }\textbf {\bibinfo {volume}
  {38}},\ \bibinfo {pages} {3098--3100} (\bibinfo {year} {1988})}\BibitemShut
  {NoStop}%
\bibitem [{\citenamefont {of~Karlsruhe}(2007)}]{TURBOMOLE}%
  \BibitemOpen
  \bibfield  {author} {\bibinfo {author} {\bibfnamefont {U.}~\bibnamefont
  {of~Karlsruhe}},\ }\href {http://www.turbomole.com} {\enquote {\bibinfo
  {title} {{TURBOMOLE V7.3 2018}, a development of {University of Karlsruhe}
  and {Forschungszentrum Karlsruhe GmbH}, {TURBOMOLE GmbH}, since 2007;},}\ }
  (\bibinfo {year} {1989-2007})\BibitemShut {NoStop}%
\bibitem [{\citenamefont {Schäfer}\ \emph {et~al.}(2000)\citenamefont
  {Schäfer}, \citenamefont {Klamt}, \citenamefont {Sattel}, \citenamefont
  {Lohrenz},\ and\ \citenamefont {Eckert}}]{COSMOiiTURBOMOLE}%
  \BibitemOpen
  \bibfield  {author} {\bibinfo {author} {\bibfnamefont {A.}~\bibnamefont
  {Schäfer}}, \bibinfo {author} {\bibfnamefont {A.}~\bibnamefont {Klamt}},
  \bibinfo {author} {\bibfnamefont {D.}~\bibnamefont {Sattel}}, \bibinfo
  {author} {\bibfnamefont {J.~C.~W.}\ \bibnamefont {Lohrenz}},\ and\ \bibinfo
  {author} {\bibfnamefont {F.}~\bibnamefont {Eckert}},\ }\bibfield  {title}
  {\enquote {\bibinfo {title} {{COSMO} implementation in {TURBOMOLE}: Extension
  of an efficient quantum chemical code towards liquid systems},}\ }\href
  {https://doi.org/10.1039/b000184h} {\bibfield  {journal} {\bibinfo  {journal}
  {Phys Chem Chem Phys}\ }\textbf {\bibinfo {volume} {2}},\ \bibinfo {pages}
  {2187--2193} (\bibinfo {year} {2000})}\BibitemShut {NoStop}%
\bibitem [{\citenamefont {Kleinschmidt}, \citenamefont {Tatchen},\ and\
  \citenamefont {Marian}(2002)}]{socoDFTMRCIwf}%
  \BibitemOpen
  \bibfield  {author} {\bibinfo {author} {\bibfnamefont {M.}~\bibnamefont
  {Kleinschmidt}}, \bibinfo {author} {\bibfnamefont {J.}~\bibnamefont
  {Tatchen}},\ and\ \bibinfo {author} {\bibfnamefont {C.~M.}\ \bibnamefont
  {Marian}},\ }\bibfield  {title} {\enquote {\bibinfo {title} {Spin-orbit
  coupling of {DFT}/{MRCI} wavefunctions: Method, test calculations, and
  application to thiophene},}\ }\href {https://doi.org/10.1002/jcc.10064}
  {\bibfield  {journal} {\bibinfo  {journal} {J Comput Chem}\ }\textbf
  {\bibinfo {volume} {23}},\ \bibinfo {pages} {824--833} (\bibinfo {year}
  {2002})}\BibitemShut {NoStop}%
\bibitem [{\citenamefont {Kleinschmidt}\ and\ \citenamefont
  {Marian}(2005)}]{egomefoesoo}%
  \BibitemOpen
  \bibfield  {author} {\bibinfo {author} {\bibfnamefont {M.}~\bibnamefont
  {Kleinschmidt}}\ and\ \bibinfo {author} {\bibfnamefont {C.~M.}\ \bibnamefont
  {Marian}},\ }\bibfield  {title} {\enquote {\bibinfo {title} {Efficient
  generation of matrix elements for one-electron spin{\textendash}orbit
  operators},}\ }\href {https://doi.org/10.1016/j.chemphys.2004.10.025}
  {\bibfield  {journal} {\bibinfo  {journal} {Chem Phys}\ }\textbf {\bibinfo
  {volume} {311}},\ \bibinfo {pages} {71--79} (\bibinfo {year}
  {2005})}\BibitemShut {NoStop}%
\bibitem [{\citenamefont {Kleinschmidt}, \citenamefont {Tatchen},\ and\
  \citenamefont {Marian}(2006)}]{SPOCKCI}%
  \BibitemOpen
  \bibfield  {author} {\bibinfo {author} {\bibfnamefont {M.}~\bibnamefont
  {Kleinschmidt}}, \bibinfo {author} {\bibfnamefont {J.}~\bibnamefont
  {Tatchen}},\ and\ \bibinfo {author} {\bibfnamefont {C.~M.}\ \bibnamefont
  {Marian}},\ }\bibfield  {title} {\enquote {\bibinfo {title} {{SPOCK}.{CI}: A
  multireference spin-orbit configuration interaction method for large
  molecules},}\ }\href {https://doi.org/10.1063/1.2173246} {\bibfield
  {journal} {\bibinfo  {journal} {J Chem Phys}\ }\textbf {\bibinfo {volume}
  {124}},\ \bibinfo {pages} {124101} (\bibinfo {year} {2006})}\BibitemShut
  {NoStop}%
\bibitem [{\citenamefont {Cambi}\ \emph {et~al.}(1991)\citenamefont {Cambi},
  \citenamefont {Cappelletti}, \citenamefont {Liuti},\ and\ \citenamefont
  {Pirani}}]{gcitopfvdwippc}%
  \BibitemOpen
  \bibfield  {author} {\bibinfo {author} {\bibfnamefont {R.}~\bibnamefont
  {Cambi}}, \bibinfo {author} {\bibfnamefont {D.}~\bibnamefont {Cappelletti}},
  \bibinfo {author} {\bibfnamefont {G.}~\bibnamefont {Liuti}},\ and\ \bibinfo
  {author} {\bibfnamefont {F.}~\bibnamefont {Pirani}},\ }\bibfield  {title}
  {\enquote {\bibinfo {title} {Generalized correlations in terms of
  polarizability for van der waals interaction potential parameter
  calculations},}\ }\href {https://doi.org/10.1063/1.461035} {\bibfield
  {journal} {\bibinfo  {journal} {The Journal of Chemical Physics}\ }\textbf
  {\bibinfo {volume} {95}},\ \bibinfo {pages} {1852--1861} (\bibinfo {year}
  {1991})}\BibitemShut {NoStop}%
\bibitem [{\citenamefont {Sharipov}\ \emph {et~al.}(2014)\citenamefont
  {Sharipov}, \citenamefont {Loukhovitski}, \citenamefont {Tsai},\ and\
  \citenamefont {Starik}}]{teodcoal2o3ncidbg}%
  \BibitemOpen
  \bibfield  {author} {\bibinfo {author} {\bibfnamefont {A.~S.}\ \bibnamefont
  {Sharipov}}, \bibinfo {author} {\bibfnamefont {B.~I.}\ \bibnamefont
  {Loukhovitski}}, \bibinfo {author} {\bibfnamefont {C.-J.}\ \bibnamefont
  {Tsai}},\ and\ \bibinfo {author} {\bibfnamefont {A.~M.}\ \bibnamefont
  {Starik}},\ }\bibfield  {title} {\enquote {\bibinfo {title} {Theoretical
  evaluation of diffusion coefficients of (al2o3)n clusters in different bath
  gases},}\ }\href {https://doi.org/10.1140/epjd/e2014-40831-2} {\bibfield
  {journal} {\bibinfo  {journal} {The European Physical Journal D}\ }\textbf
  {\bibinfo {volume} {68}} (\bibinfo {year} {2014}),\
  10.1140/epjd/e2014-40831-2}\BibitemShut {NoStop}%
\bibitem [{\citenamefont {Zhurko}\ and\ \citenamefont
  {Zhurko.}(2018)}]{ChemCraft}%
  \BibitemOpen
  \bibfield  {author} {\bibinfo {author} {\bibfnamefont {G.}~\bibnamefont
  {Zhurko}}\ and\ \bibinfo {author} {\bibfnamefont {D.}~\bibnamefont
  {Zhurko.}},\ }\href {http://www.chemcraftprog.com} {\enquote {\bibinfo
  {title} {Chemcraft. version 1.7 (build 132).}}\ } (\bibinfo {year}
  {2018})\BibitemShut {NoStop}%
\bibitem [{\citenamefont {Yang}\ and\ \citenamefont
  {Jang}(2020)}]{tionfetmipdidpbatdas}%
  \BibitemOpen
  \bibfield  {author} {\bibinfo {author} {\bibfnamefont {L.}~\bibnamefont
  {Yang}}\ and\ \bibinfo {author} {\bibfnamefont {S.~J.}\ \bibnamefont
  {Jang}},\ }\bibfield  {title} {\enquote {\bibinfo {title} {Theoretical
  investigation of non-förster exciton transfer mechanisms in perylene diimide
  donor, phenylene bridge, and terrylene diimide acceptor systems},}\ }\href
  {https://doi.org/10.1063/5.0023709} {\bibfield  {journal} {\bibinfo
  {journal} {J Chem Phys}\ }\textbf {\bibinfo {volume} {153}},\ \bibinfo
  {pages} {144305} (\bibinfo {year} {2020})}\BibitemShut {NoStop}%
\bibitem [{\citenamefont {Duschinsky}(1937)}]{tiotesimamctFCp}%
  \BibitemOpen
  \bibfield  {author} {\bibinfo {author} {\bibfnamefont {F.}~\bibnamefont
  {Duschinsky}},\ }\bibfield  {title} {\enquote {\bibinfo {title} {The
  importance of the electron spectrum in multi atomic molecules concerning the
  franck-condon principle},}\ }\href@noop {} {\bibfield  {journal} {\bibinfo
  {journal} {Acta Physicochim. U.R.S.S.}\ }\textbf {\bibinfo {volume} {7}},\
  \bibinfo {pages} {551--566} (\bibinfo {year} {1937})}\BibitemShut {NoStop}%
\bibitem [{\citenamefont {Burgdorff}, \citenamefont {Ehrhardt},\ and\
  \citenamefont {Loehmannsroeben}(1991)}]{ppotdis2htd}%
  \BibitemOpen
  \bibfield  {author} {\bibinfo {author} {\bibfnamefont {C.}~\bibnamefont
  {Burgdorff}}, \bibinfo {author} {\bibfnamefont {S.}~\bibnamefont
  {Ehrhardt}},\ and\ \bibinfo {author} {\bibfnamefont {H.~G.}\ \bibnamefont
  {Loehmannsroeben}},\ }\bibfield  {title} {\enquote {\bibinfo {title}
  {Photophysical properties of tetracene derivatives in solution. 2.
  halogenated tetracene derivatives},}\ }\href
  {https://doi.org/10.1021/j100164a016} {\bibfield  {journal} {\bibinfo
  {journal} {The Journal of Physical Chemistry}\ }\textbf {\bibinfo {volume}
  {95}},\ \bibinfo {pages} {4246--4249} (\bibinfo {year} {1991})}\BibitemShut
  {NoStop}%
\bibitem [{\citenamefont {Lakowitz}(2007)}]{lakowitz}%
  \BibitemOpen
  \bibfield  {author} {\bibinfo {author} {\bibfnamefont {J.~R.}\ \bibnamefont
  {Lakowitz}},\ }\href
  {https://www.ebook.de/de/product/19111410/joseph_r_lakowicz_principles_of_fluorescence_spectroscopy.html}
  {\emph {\bibinfo {title} {Principles of Fluorescence Spectroscopy}}},\
  \bibinfo {edition} {3rd}\ ed.\ (\bibinfo  {publisher} {Springer US},\
  \bibinfo {year} {2007})\BibitemShut {NoStop}%
\bibitem [{\citenamefont {Akselrod}\ \emph
  {et~al.}(2014{\natexlab{b}})\citenamefont {Akselrod}, \citenamefont
  {Deotare}, \citenamefont {Thompson}, \citenamefont {Lee}, \citenamefont
  {Tisdale}, \citenamefont {Baldo}, \citenamefont {Menon},\ and\ \citenamefont
  {Bulovic}}]{Akselrod2014}%
  \BibitemOpen
  \bibfield  {author} {\bibinfo {author} {\bibfnamefont {G.~M.}\ \bibnamefont
  {Akselrod}}, \bibinfo {author} {\bibfnamefont {P.~B.}\ \bibnamefont
  {Deotare}}, \bibinfo {author} {\bibfnamefont {N.~J.}\ \bibnamefont
  {Thompson}}, \bibinfo {author} {\bibfnamefont {J.}~\bibnamefont {Lee}},
  \bibinfo {author} {\bibfnamefont {W.~A.}\ \bibnamefont {Tisdale}}, \bibinfo
  {author} {\bibfnamefont {M.~A.}\ \bibnamefont {Baldo}}, \bibinfo {author}
  {\bibfnamefont {V.~M.}\ \bibnamefont {Menon}},\ and\ \bibinfo {author}
  {\bibfnamefont {V.}~\bibnamefont {Bulovic}},\ }\bibfield  {title} {\enquote
  {\bibinfo {title} {{Visualization of exciton transport in ordered and
  disordered molecular solids}},}\ }\href {https://doi.org/10.1038/ncomms4646}
  {\bibfield  {journal} {\bibinfo  {journal} {Nature Communications}\ }\textbf
  {\bibinfo {volume} {5}},\ \bibinfo {pages} {1--8} (\bibinfo {year}
  {2014}{\natexlab{b}})}\BibitemShut {NoStop}%
\bibitem [{\citenamefont {Senes}\ \emph {et~al.}(2016)\citenamefont {Senes},
  \citenamefont {Meskers}, \citenamefont {Dijkstra}, \citenamefont {van
  Franeker}, \citenamefont {Altazin}, \citenamefont {Wilson},\ and\
  \citenamefont {Janssen}}]{tdmoifospfoitioma}%
  \BibitemOpen
  \bibfield  {author} {\bibinfo {author} {\bibfnamefont {A.}~\bibnamefont
  {Senes}}, \bibinfo {author} {\bibfnamefont {S.~C.~J.}\ \bibnamefont
  {Meskers}}, \bibinfo {author} {\bibfnamefont {W.~M.}\ \bibnamefont
  {Dijkstra}}, \bibinfo {author} {\bibfnamefont {J.~J.}\ \bibnamefont {van
  Franeker}}, \bibinfo {author} {\bibfnamefont {S.}~\bibnamefont {Altazin}},
  \bibinfo {author} {\bibfnamefont {J.~S.}\ \bibnamefont {Wilson}},\ and\
  \bibinfo {author} {\bibfnamefont {R.~A.~J.}\ \bibnamefont {Janssen}},\
  }\bibfield  {title} {\enquote {\bibinfo {title} {Transition dipole moment
  orientation in films of solution processed fluorescent oligomers:
  investigating the influence of molecular anisotropy},}\ }\href
  {https://doi.org/10.1039/c5tc03481g} {\bibfield  {journal} {\bibinfo
  {journal} {Journal of Materials Chemistry C}\ }\textbf {\bibinfo {volume}
  {4}},\ \bibinfo {pages} {6302--6308} (\bibinfo {year} {2016})}\BibitemShut
  {NoStop}%
\end{thebibliography}%

\end{document}